\documentclass[%
 reprint,
superscriptaddress,
 amsmath,amssymb,
 aps,
]{revtex4-2}

\usepackage[utf8]{inputenc}
\usepackage{graphicx}
\usepackage{color}
\usepackage{physics}
\usepackage{braket}
\usepackage{comment}
\usepackage[version=4]{mhchem}
\usepackage[breaklinks=true]{hyperref}
\newcommand{\mr}[1]{\mathrm{#1}}
\date{\today}
\usepackage{amsthm}
\theoremstyle{definition}
\newtheorem{theorem}{Theorem}[]

\newtheorem{proposition}[theorem]{Proposition}
\newtheorem*{proposition2*}{Proposition}

\begin{document}
\title{Deep variational quantum eigensolver for excited states and its application to quantum chemistry calculation of periodic materials}

\author{Kaoru Mizuta}
\email{mizuta.kaoru.65u@st.kyoto-u.ac.jp}
\affiliation{QunaSys Inc., Aqua Hakusan Building 9F, 1-13-7 Hakusan, Bunkyo, Tokyo 113-0001, Japan}
\affiliation{Department of Physics, Kyoto University, Kyoto 606-8502, Japan}

\author{Mikiya Fujii}
\affiliation{Technology division, Innovation Promotion Sector, Panasonic Corporation, 1006 Kadoma, Kadoma City, Osaka 571-8508, Japan}

\author{Shigeki Fujii}
\affiliation{Technology division, Innovation Promotion Sector, Panasonic Corporation, 1006 Kadoma, Kadoma City, Osaka 571-8508, Japan}

\author{Kazuhide Ichikawa}
\affiliation{Technology division, Innovation Promotion Sector, Panasonic Corporation, 1006 Kadoma, Kadoma City, Osaka 571-8508, Japan}

\author{Yutaka Imamura}
\affiliation{Innovative Technology Laboratories, AGC Inc., Yokohama 230-0045, Japan}

\author{Yukihiro Okuno}
\affiliation{Analysis Technology Center, FUJIFILM Corporation, Minamiashigara City, Kanagawa 250-0193, Japan}

\author{Yuya O. Nakagawa}
\affiliation{QunaSys Inc., Aqua Hakusan Building 9F, 1-13-7 Hakusan, Bunkyo, Tokyo 113-0001, Japan}

\begin{abstract}
A programmable quantum device that has a large number of qubits without fault-tolerance has emerged recently.
Variational Quantum Eigensolver (VQE) is one of the most promising ways to utilize the computational power of such devices to solve problems in condensed matter physics and quantum chemistry.
As the size of the current quantum devices is still not large for rivaling classical computers at solving practical problems, Fujii \textit{et al.} proposed a method called ``Deep VQE" which can provide the ground state of a given quantum system with the smaller number of qubits by combining the VQE and the technique of coarse-graining~\cite{fujii2020deep}.
In this paper, we extend the original proposal of Deep VQE to obtain the excited states and apply it to quantum chemistry calculation of a periodic material, which is one of the most impactful applications of the VQE.
We first propose a modified scheme to construct quantum states for coarse-graining in Deep VQE to obtain the excited states.
We also present a method to avoid a problem of meaningless eigenvalues in the original Deep VQE without restricting variational quantum states. 
Finally, we classically simulate our modified Deep VQE for quantum chemistry calculation of a periodic hydrogen chain as a typical periodic material.
Our method reproduces the ground-state energy and the first-excited-state energy with the errors up to $O(1)\%$ despite the decrease in the number of qubits required for the calculation by two or four compared with the naive VQE.
Our result will serve as a beacon for tackling quantum chemistry problems with classically-intractable sizes by smaller quantum devices in the near future.
\end{abstract}

\maketitle

\section{\label{sec:intro} Introduction}
Noisy Intermediate-Scale Quantum (NISQ) devices have a moderate number [$O(10)-O(100)$] of qubits that we can control very precisely  although they are not fault-tolerant~\cite{Preskill2018, moll2018quantum}.
Variational Quantum Eigensolver (VQE), which computes the approximate ground state of quantum systems, is one of the most promising applications of the NISQ devices~\cite{peruzzo2014variational}.
As the size and hardware-precision of the NISQ devices have been  growing recently~\cite{Arute2020,Zhong2020quantum}, extensive efforts are put into studies of the VQE, which leads to various extension of the VQE to its practical application to condensed matter physics and quantum chemistry.
For instance, while the original VQE gives the approximate ground state, a bunch of methods to calculate low-energy excited states \cite{McClean2017hybrid,nakanishi2018subspace,Parrish2019,higgott2019variational,jones2019variational, Ollitrault2020quantum} were proposed.
There are also VQE-based algorithms for calculating Green's functions 
\cite{endo2020calculation, rungger2020dynamical}, nonequilibrium dynamics~\cite{li2017efficient, yuan2019theory,endo2020variational}, and its steady states~\cite{yoshioka2020nonequilibrium} in dissipative systems, etc.
As for applications to quantum chemistry~\cite{mcardle2018quantum,Cao2018}, energies of electronic states of molecules were already computed experimentally by using the VQE~\cite{peruzzo2014variational,kandala2017hardware,colless2018computation,kandala2019error}.
Methods to obtain other important quantities such as the energy derivatives~\cite{Mitarai2020theory, Obrien2019} and the non-adiabatic coupling~\cite{tamiya2020calculating} were proposed.
Moreover, the possibility to explore periodic materials by the VQE was examined in Refs.~\cite{liu2020simulating,manrique2020momentumspace,yoshioka2020periodic}.

While the above-mentioned methods based on the VQE will be utilized in the near future, the currently-available NISQ devices do not reach the stage to overwhelm computations by classical computers because the number of qubits and the precision of gate-operations on the qubits are still limited.
To relax the requirement for the hardware of the NISQ devices, K. Fujii \textit{et al.} recently proposed a method called Deep Variational Quantum Eigensolver (Deep VQE)~\cite{fujii2020deep}.
Deep VQE combines the VQE and an idea of the divide-and-conquer method that is popular in quantum chemistry~\cite{Yang1995,Gordon2012,yamazaki2018practical, kawashima2021efficient}.
By performing the coarse-graining of an original large problem based on the solutions of the VQE in smaller subsystems,
Deep VQE allows us to obtain the approximate ground state of the original problem with the smaller number of qubits compared to the usual VQE.

However, from both physical and technical points of view, it has been still nontrivial and essential whether Deep VQE can deal with low-energy properties, including excited states, which are of great interest both in physics and chemistry.
The first problem for treating excited states is that coarse-graining in the original Deep VQE aims to capture only the ground state by considering excitations from inter-subsystem interactions but neglecting those from intra-subsystem interactions.
We should carefully consider whether Deep VQE can yield low-energy excited states from the physical point of view, e.g., the type of excitations to be included in the algorithm.
The second problem is a technical one that imposes strong limitations on variational quantum states employed in Deep VQE; that is, if we choose variational quantum states freely, wrong eigenvalues may be obtained because of the appearance of meaningless eigenvalues in the energy spectrum.
This problem occurs not only for excited states but also for the ground state, and the use of special variational quantum states to avoid the meaningless eigenvalues is mentioned in the original Deep VQE proposal~\cite{fujii2020deep}.

The first purpose of this paper is to elaborate the protocol of Deep VQE for correctly simulating low-energy eigenstates by solving the above-mentioned problems.
Concretely, we propose a local basis set in which both excitations from intra-subsystem and inter-subsystem interactions are taken into account in the procedure of the coarse-graining.
We show that our modified choice of the local basis set enables us to obtain more accurate low-energy eigenvalues than those obtained with the original Deep VQE by employing the perturbation theory and the Quantum Subspace Expansion (QSE) method~\cite{McClean2017hybrid}.
We also address the technical problem by giving an elaborate construction of a coarse-grained Hamiltonian with additional penalty terms.
The penalty terms make meaningless eigenvalues move away from the low-energy spectrum, enabling us to use any kind of variational quantum states. We provide a mathematically rigorous sufficient condition for the magnitude of the penalty terms to realize arbitrary options of variational quantum states. 
Finally, we numerically confirm those findings by classical simulations in a spin chain.

The second purpose is to examine the validity of Deep VQE for quantum chemistry problems, which has been raised as one of the motivations of the original Deep VQE \cite{fujii2020deep}.
In particular, we focus on low-energy eigenstates of periodic materials.
Performing quantum chemistry calculations for periodic materials remains one of the ultimate goals of quantum chemistry and material science, but the huge computational cost hinders its realization.
The use of quantum computers (possibly NISQ devices)~\cite{liu2020simulating,manrique2020momentumspace,yoshioka2020periodic} may circumvent the situation, but it still requires the large number of qubits to perform the calculation. 
In this regards, the coarse-graining techniques such as Deep VQE are of great importance for periodic materials. 
We take a periodic hydrogen chain as the most straightforward example and classically simulate the performance of our modified Deep VQE by calculating its ground-state energy and its first-excited-state energy.
We find a proper way of the coarse-graining and obtain the low-energy eigenvalues with an error up to $O(1)\%$ despite reducing the number of qubits by up to four.
Our results will serve as a beacon for simulating low-energy properties of a variety of materials with smaller quantum devices in the coming NISQ era.

This paper is organized as follows.
In Sec.~\ref{sec:preli}, we review Deep VQE and its pros and cons to describe our study's motivation in detail.
Sections~\ref{sec:excited} and \ref{sec:chemistry} provide the main results of our research.
In Sec.~\ref{sec:excited}, we formulate Deep VQE for low-energy excited states and show numerical results for simple spin systems. We discuss the validity of our proposed method with relating to other methods such as QSE.
In Sec.~\ref{sec:chemistry}, we apply our modified Deep VQE to a periodic hydrogen chain and show numerical results for the ground-state energy the first-excited-state energy. 
We conclude this paper in Section~\ref{sec:conclusions}.

\section{\label{sec:preli} Preliminaries}
In this section, we review the algorithm of the original Deep VQE proposed in Ref.~\cite{fujii2020deep} and discuss its pros and cons to motivate our study.
Deep VQE combines the VQE~\cite{peruzzo2014variational} and the divide-and-conquer method~\cite{Yang1995,Gordon2012}, with which we can find the ground state with fewer qubits than those required when using the usual VQE.

\subsection{\label{subsec:deepvqe protocol} Protocol of Deep VQE}
\begin{figure}
\begin{center}
    \includegraphics[height=5cm, width=9cm]{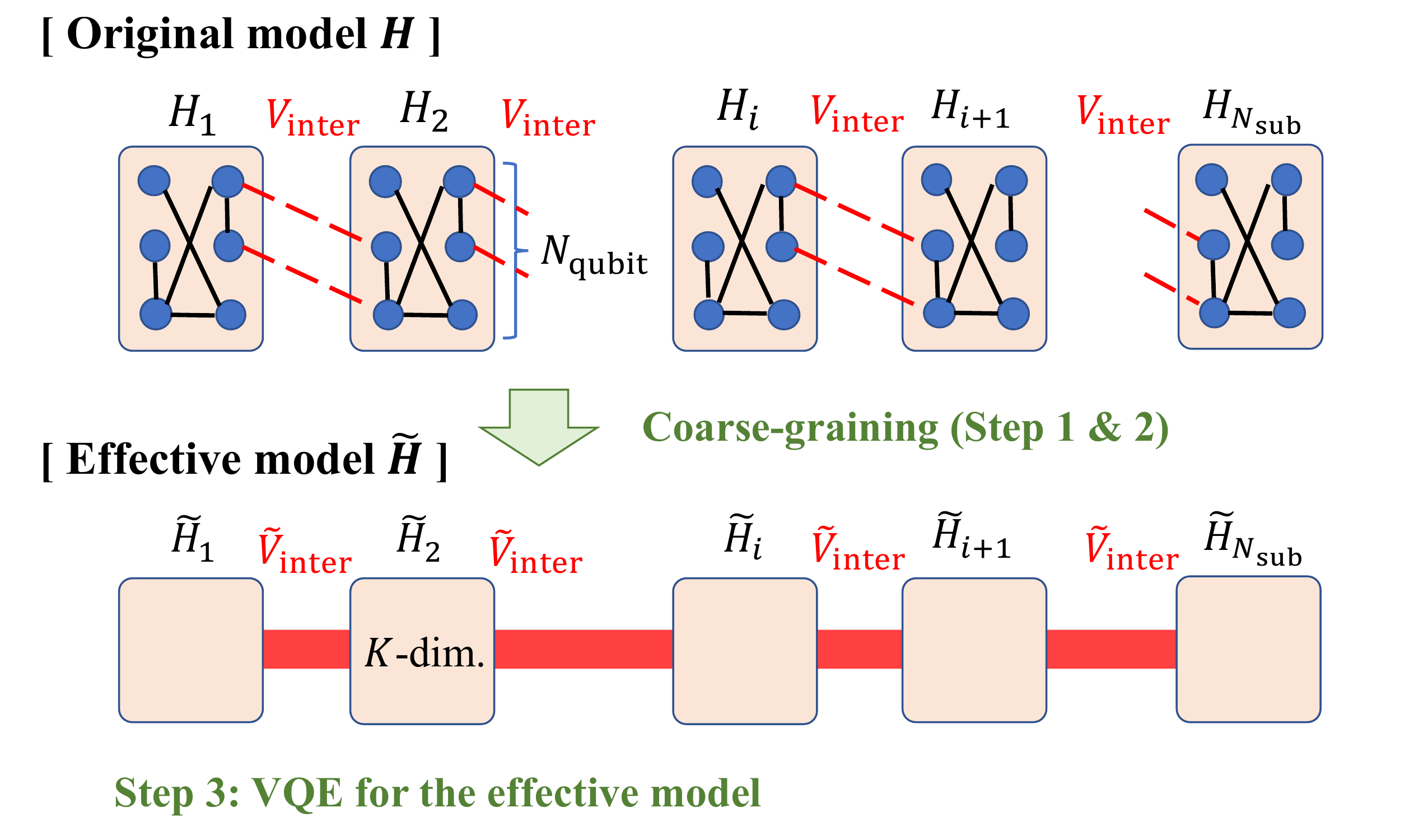}
    \caption{Schematic picture of Deep VQE. The details are described in Sec.~\ref{subsec:deepvqe protocol}.} 
    \label{fig:deepvqe}
 \end{center}
 \end{figure}

We consider a $N_\mr{tot}$-qubit quantum system and split it into $N_{\text{sub}}$ subsystems, where each subsystem is composed of $N_{\text{qubit}}=N_\mr{tot}/N_{\text{sub}}$ qubits (see Fig. \ref{fig:deepvqe}). Then, we assume that the Hamiltonian of the system is written in the following form:
\begin{equation}\label{eq: hamiltonian}
H = \sum_{i=1}^{N_{\text{sub}}} H_i + V_{\text{inter}}, \quad V_{\text{inter}} = \sum_{i \neq j}^{N_{\text{sub}}} \sum_\alpha \nu_{ij}^\alpha V_i^\alpha \otimes W_j^\alpha,
\end{equation}
where $H_i$ represents an intra-subsystem Hamiltonian of the $i$-th subsystem. The inter-subsystem interactions are described by $V_{\text{inter}}$, in which $V_i^\alpha$ and $W_j^\alpha$ are operators acting on the $i$-th and the $j$-th subsystems, respectively. The index $\alpha$ enumerates such interaction terms between the subsystems $i$ and $j$.
The above equation only includes inter-subsystem interactions over two subsystems, but the extension to the cases where inter-subsystem interactions involves more than three subsystems is straightforward. Under this setup, Deep VQE is composed of the following three steps.

\begin{flushleft}
\underbar{Step 1}: Perform VQE for each subsystem.
\end{flushleft}

We obtain a set of local ground states $\{ \ket{\psi_0}_i \}_{i=1}^{N_{\text{sub}}}$, where $\ket{\psi_0}_i$ represents the ground state of the $i$-th subsystem Hamiltonian $H_i$, by the ordinary VQE. This can be completed by minimizing the cost function $\bra{\psi (\vec{\theta})} H_i \ket{ \psi (\vec{\theta})}$ with a variational quantum state $\ket{\psi(\theta)} = U(\vec{\theta}) \ket{0}$ and setting $\ket{\psi_0}_i = \ket{\psi(\vec{\theta}_i^\ast)}$, where $\vec{\theta}_i^\ast$ is the optimal parameter set for $i$-th subsystem. This step requires a $N_{\text{qubit}}$-qubit quantum computer.

\begin{flushleft}
\underbar{Step 2}: Construct an effective model in the restricted Hilbert space (coarse-graining).
\end{flushleft}

We use a divide-and-conquer method here. For each subsystem, we choose a set of local excitation operators $\{ P_k^{(i)} \}_{k=1}^{K}$ with $P_1^{(i)}=I$ (identity operator). Let $\tilde{\mathcal{H}}_i$ denote the  $K$-dimensional local Hilbert space spanned by $\{ P_k^{(i)} \ket{\psi_0}_i \}_{k=1}^{K}$. We focus on the restricted Hilbert space $\tilde{\mathcal{H}} = \otimes_i^{N_\text{sub}} \tilde{\mathcal{H}}_i$, and construct the effective Hamiltonian after coarse-graining by
\begin{equation}\label{eq: hamiltoniantilde}
\tilde{H} = \left. H \right|_{\tilde{\mathcal{H}} } = \sum_{i=1}^{N_{\text{sub}}} \tilde{H}_i + \sum_{i \neq j}^{N_{\text{sub}}} \sum_\alpha \nu_{ij}^\alpha \tilde{V}_i^\alpha \otimes \tilde{W}_j^\alpha,
\end{equation}
which is reminiscent of the internally contracted multireference-configuration interaction method in quantum chemistry \cite{Werner1982,Werner1988}. Here, we define  $\tilde{H}_i=\left. H_i \right|_{\tilde{\mathcal{H}_i} }$, $\tilde{V}_i^\alpha=\left. V_i^\alpha \right|_{\tilde{\mathcal{H}_i} }$, and $\tilde{W}_j^\alpha=\left. W_j^\alpha \right|_{\tilde{\mathcal{H}_j} }$.
We should compute $K \times K$ matrix representations of $\tilde{H}_i$, $\tilde{V}_i^\alpha$ and $\tilde{W}_j^\alpha$ by using the result of Step 1 and the $N_\text{qubit}$-qubit quantum computer.
Recalling that the local basis $\{ P_k^{(i)} \ket{\psi_0}_i \}_{k=1}^K$ is not orthonormal, we first construct an orthonormal basis $\{ \ket{{\phi}_k}_i \}_{k=1}^K$ by the Gram-Schmidt method. Concretely, we compute a $K\times K$ matrix $G^{(i)}$ representing the inner product,
\begin{equation}\label{eq:overlap}
G^{(i)}_{kl} =  \,_i\! \bra{\psi_0} P_k^{(i) \dagger} P_l^{(i)} \ket{\psi_0}_i, \quad k,l = 1,2, \hdots , K,
\end{equation}
by the $N_{\text{qubit}}$-qubit quantum device.
With classical computation using the matrix $G^{(i)}$, we can obtain a $K \times K$ matrix $S^{(i)}$ so that the set of states, given by
\begin{equation}
\ket{{\phi}_k}_i = \sum_{k^\prime=1}^K S_{k^\prime k}^{(i)} (P_{k^\prime}^{(i)} \ket{{\psi}_0}_i ), \quad k=1,2,\hdots, K,
\end{equation}
can provide an orthonormal basis of $\tilde{\mathcal{H}}_i$. We then obtain the matrix elements
\begin{eqnarray}
(\tilde{A})_{kl} &=& \,_i\! \bra{\phi_k} A \ket{\phi_l}_i \nonumber \\
&=& \sum_{k^\prime, l^\prime} S_{k^\prime k}^{(i) \ast} S_{l^\prime l}^{(i)} \,_i\! \bra{\psi_0} P_{k^\prime}^{(i),\dagger} A P_{l^\prime}^{(i)} \ket{\psi_0}_i \label{eq:matrix_elements}
\end{eqnarray}
by using the $N_\text{qubit}$-qubit quantum device of to evaluate the expectation values of $P_{k^\prime}^{(i),\dagger} A P_{l^\prime}^{(i)}$ for $A = H_i, V_i^\alpha \text{ or } W_i^\alpha$.

In this step, there are several options for the choice of the local excitation operators $\{ P_k^{(i)} \}$, or equivalently the local basis set $\{ \ket{\phi_k}_i \}$.
In Ref.~\cite{fujii2020deep}, a series of local operators that occur in the inter-subsystem interactions and nontrivially act on the boundary sites of subsystems are adopted as $\{ P_k^{(i)} \}$ for evaluating the ground state of spin systems.

\begin{flushleft}
\underbar{Step 3}: Perform VQE for the effective model.
\end{flushleft}

We finally compute the ground state of the effective Hamiltonian $\tilde{H}$ by using VQE with expecting that it reproduces the essence of the ground state of the original Hamiltonian $H$.
To perform VQE for $\tilde{H}$ on quantum computers composed of qubits, we should choose an integer $N_\text{eff} = \lceil \log_2 K \rceil$ to embed $\tilde{H}$ into a Hilbert space of $N_\text{eff}$ qubits.
We add auxiliary dimensions to each local Hilbert space $\tilde{\mathcal{H}}_i$  and replace the effective Hamiltonian by
\begin{eqnarray}
\tilde{H}_{\text{eff}} &=& \sum_{i=1}^{N_{\text{sub}}} \tilde{H}_{i,\text{eff}} + \sum_{i \neq j}^{N_{\text{sub}}} \sum_\alpha  \nu_{ij}^\alpha \tilde{V}_{i,\text{eff}}^\alpha \otimes \tilde{W}_{j,\text{eff}}^\alpha,  \label{eq: hamiltonian_eff_1} \\
\tilde{H}_{i,\text{eff}} &=& \tilde{H}_{i} \oplus 0_{2^{N_{\text{eff}}}-K}, \label{eq: hamiltonian_eff_2} \\ \tilde{V}_{i,\text{eff}}^\alpha &=& \tilde{V}_{i}^\alpha \oplus 0_{2^{N_{\text{eff}}}-K}, \quad \tilde{W}_{j,\text{eff}}^\alpha = \tilde{W}_{j}^\alpha \oplus 0_{2^{N_{\text{eff}}}-K}, \label{eq: hamiltonian_eff_3}
\end{eqnarray}
where $0_M$ represents a $M \times M$ zero matrix. The replaced effective Hamiltonian $\tilde{H}_{\text{eff}}$ is defined on $N_{\text{sub}} \times N_{\text{eff}}$ qubits, and we can perform VQE for it.
By carefully choosing a variational quantum circuit $V(\vec{\theta})$ so that $V(\vec{\theta}) \ket{0}$ has no components out of the $K^{N_{\text{sub}}}$-dimensional subspace corresponding to $\tilde{\mathcal{H}}$, we obtain the ground-state energy by minimizing the cost function $\bra{0} V^\dagger (\vec{\theta}) \tilde{H}_\mr{eff} V(\vec{\theta}) \ket{0}$.
This Step 3 requires quantum computation on $(N_{\text{sub}} \times N_{\text{eff}}) = N_{\text{sub}} \lceil \log_2 K \rceil$ qubits.

Steps 1,2 and 3 are crucial procedures of Deep VQE.
The number of qubits required for the whole algorithm is given by
\begin{equation}\label{eq:N_required}
    N_\mr{req} = \mr{max} \left( N_\mr{qubit}, \, N_\mr{sub} \times N_\mr{eff} \right).
\end{equation}
When $K$ is not taken as so large [$\sim O(e^{N_\text{qubit}})$], $N_\mr{req}$ is small compared to the required number of qubits for simulating the original system, $N_\text{tot} = N_\text{sub} N_\text{qubit}$.
In the original proposal~\cite{fujii2020deep}, the repetition of Steps 1-3 are also proposed to further reduce the number of qubits.

\subsection{\label{subsec:pros_and_cons} Motivation of our study}
The pros of Deep VQE is that we can calculate the ground state with the smaller number of qubits. The key to reducing the required qubits is the choice of local excitation operators $\{ P_k^{(i)} \}_{k=1}^K$ in Step 2, in which we discard some information of the whole Hilbert space.
As is usual with the divide-and-conquer methods, it is argued in Ref.~\cite{fujii2020deep} that the resulting restricted Hilbert space can describe the ground state of the original Hamultonian well as long as it is low-entangled and we can choose the proper local basis $\{ P_k^{(i)} \}_{k=1}^K$.


We motivate our study by raising several points for the original Deep VQE proposal.
The first point is whether it is possible to describe the low-energy eigenstates of the original Hamiltonian in the restricted Hilbert space, or simply stating, the applicability of Deep VQE to the excited states.
It is nontrivial whether the same local basis for the ground state works also for the excited states.
The second point lies in Step 3.
When we replace the effective Hamiltonian $\tilde{H}$ by $\tilde{H}_\mr{eff}$ with inserting auxiliary dimensions, some meaningless (or artificial) eigenvalues appear in the spectrum of $\tilde{H}_\mr{eff}$.
As we explicitly show in Sec.~\ref{subsec:modified2}, this can alter both the ground-state energy and the low-excited-state energies of $\tilde{H}_\mr{eff}$ from those of $\tilde{H}$.
This is the reason why we restrict the variational quantum states in Step 3 following Ref.~\cite{fujii2020deep}.
In practice, construction of such variational quantum states is not so straightforward and it is desirable to develop a way to use arbitrary variational quantum states for searching low-energy eigenstates of $\tilde{H}_\mr{eff}$.
The third point relates to the applicability of Deep VQE to fermionic systems such as quantum chemistry problems.
The performance of Deep VQE in fermionic systems has not been explored so far, where quantum states mapped to qubit systems can be more entangled in general because of the non-local terms in the mapped Hamiltonian.

This study aims to construct the modified Deep VQE protocol for low-energy eigenstates by solving the above problems and to examine its practical use for chemistry problems, particularly the simulation of periodic materials. The simulation of periodic materials in \textit{ab initio} level, which always lies at the center of condensed matter physics, requires many qubits to predict accurate results in the thermodynamic limit.
Thus, periodic materials are expected to be a good target of the modified Deep VQE, in that we would like to obtain both their ground state and the low-energy-excited states with the smaller number of qubits.

\section{\label{sec:excited} Main Result: Deep VQE for low-energy physics}
This section provides the first half of the main results. As discussed in the previous section, the original Deep VQE has some nontrivial problems in calculating low-energy eigenstates.
In the subsections \ref{subsec:modified1} and \ref{subsec:modified2}, we propose the modified Deep VQE protocol for excited states that solves these problems. In the subsection \ref{subsec:spinchain}, we provide a numerical example to confirm the validity of our modified Deep VQE.

\subsection{\label{subsec:modified1} Choice of local basis for obtaining accurate excited-states}
Here, we provide the modified way of choosing the local basis in Step 2 of Deep VQE so that low-energy eigenstates of the original Hamiltonian can be well represented in the restricted Hilbert space.

Before proposing our modified choice of the local basis to obtain accurate excited states, let us recall the local basis of the original Deep VQE and argue that it may fail for the excited states.
The local basis $\{ P_k^{(i)} \}_k$ in the original Deep VQE is chosen as a set of Pauli operators contained in the inter-subsystem interactions (an explicit example can be found in Sec.~\ref{subsec:spinchain}). 
Assuming that $H_i$ has a unique ground state $\ket{\psi_0}_i$ and that the VQE in Step 1 of Deep VQE is accurate enough, it follows that $\ket{\Psi_0}=\bigotimes_i \ket{\psi_0}_i$ is the ground state of $H_\mr{intra}=\sum_i H_i$.
Regarding $H_\mr{intra}$ and $V_\mr{inter}$ as an unperturbed Hamiltonian and a perturbation, respectively, Deep VQE always gives the ground-state energy more accurate than that of the first-order perturbation theory, $\braket{\Psi_0 | H | \Psi_0}$.
This is because the restricted Hilbert space $\tilde{\mathcal{H}}$ includes $\ket{\Psi_0}$ by construction.
Although neither the original Deep VQE overwhelms higher-order perturbation theories nor vice versa, they commonly capture local excitations from the unperturbed ground state $\ket{\Psi_0}$ invoked by $\{ P_k^{(i)} \}$ (or $V_\mr{inter}$), which thereby validate the accuracy of the original Deep VQE for the ground state.
On the other hand, the restricted Hilbert space of the original Deep VQE cannot reproduce the low-energy-excited-states eigenvalues even at the first-order perturbation level.
The $n$-th excited states $\ket{\Psi_n}$ of the unperturbed Hamiltonian $H_\mr{intra}$ is not generated by applying only the excitation operators stemming from the inter-subsystem interactions to $\ket{\Psi_0}$.
Hence, the choice of the local basis of the original Deep VQE will fail to produce accurate excited-state energies (see also the numerical results in the subsection~\ref{subsec:spinchain}).

We propose an operator set $\{ P_k^{(i)} \}_{k=1}^K$ for properly representing low-energy excited states as the modified Deep VQE protocol. The main idea is to take into account not only excitations by inter-subsystem interactions $V_\mr{inter}$ but also those by intra-subsystem interactions $H_\mr{intra}$.
Namely, the simplest choice of $\{ P_k^{(i)} \}_{k=1}^K$ is
\begin{equation}\label{eq:localop_bulk}
    P_k^{(i)} \in \mathcal{W} := \{ I \} \cup \left( \bigcup_{j=1}^{N_\mr{qubit}} \{ X_j,Y_j,Z_j \} \right),
\end{equation}
composed of the identity operator and Pauli operators on each sites of the subsystem $i$.
The local Hilbert space dimension $K=3N_\mr{qubit}+1$ is still much smaller than the original one $2^{N_\mr{qubit}}$, meaning the decrease of qubits required for simulation.

We can show the validity of the modified local basis generated by $\mathcal{W}$ for Deep VQE of the excited states under certain assumptions.
Concretely, we can show that our modified Deep VQE yields more accurate low-excited-state energies than the first-order perturbation theory of $V_\mr{inter}$.
We assume that the low-lying excited states of the $i$-th subsystem, $\ket{\psi_n}_i$ ($n=1,2,\hdots$), can be well described by the restricted Hilbert space for the subsystem, $\{  P_k^{(i)}\ket{\psi_0}_i | P_k^{(i)} \in \mathcal{W} \}_{k=1}^K$.
This assumption is justified when the excited states of the subsystem are not so entangled and approximately generated only from the first-order excitation of the terms included in $H_i$.
This assumption can also be rephrased that the QSE method~\cite{McClean2017hybrid} works well in the $i$-th subsystem, where a similar restricted Hilbert space is constructed and the Hamiltonian within that space is solved classically.
To see how the assumption leads to  the validity of our choice of the local basis, let us consider the first-excited states of the original Hamiltonian composed of $N_\mr{sub}$ identical subsystems.
The degenerated first-excited states of the unperturbed Hamiltonian $H_\mr{intra}$,
\begin{equation}\label{eq:first_state}
    \ket{\Psi_1^{(i)}} = \ket{\psi_1}_i \otimes \left( \bigotimes_{j: \, j \neq i}^{N_\mr{sub}} \ket{\psi_0}_j \right),
\end{equation}
are included in the subspace $\tilde{\mathcal{H}}$ because of the assumption.
The perturbation theory for $V_\mr{inter}$ tells us that the subspace $\mathcal{S}_1=\mr{span} (\{ \ket{\Psi_1^{(i)}} | \, i=1,\hdots,N_\mr{sub} \})$ can reproduce the approximate first-excited-state energy as
\begin{equation}
   E_1 (\mr{local}) = \min \left\{ \left. \frac{\braket{\Psi|H|\Psi}}{\braket{\Psi|\Psi}} \, \right| \, \ket{\Psi} \in \mathcal{S}_1 \right\}.
\end{equation}
Because of the inclusion $\mathcal{S}_1 \subset \tilde{\mathcal{H}}$ by construction, we can conclude that the modified Deep VQE always provides more accurate low-excited energies than that of the first-order perturbation.

The higher-order perturbation theories involve an infinite series of the unperturbed eigenstates $\ket{\Psi_k}$, thereby making it difficult to obtain rigorous relation with the Deep VQE. However, we can intuitively give some correspondence as well as the ground states. For instance, in the second-order perturbation theory without degeneracy, the unperturbed eigenstates $\ket{\Psi_k}$ with the large value of $|\braket{\Psi_k | V_\mr{inter} | \Psi_n } / (\braket{\Psi_n | H_\mr{intra} | \Psi_n }-\braket{\Psi_k | H_\mr{intra} | \Psi_k }) |$ have principal contributions to the $n$-th low-energy eigenvalues. Such states $\ket{\Psi_k}$, having the low-energy under $H_\mr{intra}$ and the connection via local excitations from $V_\mr{inter}$, are approximately included in the restricted Hilbert space of our method. Thus, our method is expected to provide approximate low-energy eigenvalues as well as the higher-order pertubation theories. In fact, in the later numerical simulations, we will see that the modified Deep VQE comparably outperforms the first-order perturbation result, supporting this expectation.

We note another choice of the local basis set to evaluate the excited states.
One possible choice is to include the states like $\{ P_k^{(i)} \ket{\psi_n}_i\}$, where $\ket{\psi_n}_i$ is the subsystem excited eigenstate.
Since the restricted Hilbert space after the coarse-graining directly includes $\ket{\Psi_1^{(i)}}$ [Eq. (\ref{eq:first_state})], we can expect that this choice also gives approximate excited states.
The caveat of choice is that we should employ a proper algorithms to calculate the excited states $\ket{\psi_n}_i$ in the subsystems to construct evaluate the matrix elements of $\tilde{H}$.
We discuss the details of this choice of the local basis in Appendix~\ref{Aseq:multi_state}.

To summarize, in the modified Deep VQE for calculating low-energy eigenstates, we prepare local excitation operators in Step 2 as $\mathcal{W}$ (Eq.~\eqref{eq:localop_bulk}).
This corresponds to taking into account excitations by both intra-subsystem and inter-subsystem terms. 

\subsection{\label{subsec:modified2} Construction of effective qubit model with penalty}

Next, we provide the modified way of constructing the effective qubit model in Step 3 of Deep VQE so that we can adopt any variational quantum circuit.

Let us first see the problem of meaningless eigenvalues in the original Deep VQE by taking simple examples.
In Step 3 of Deep VQE, we insert additional dimensions to the effective Hamiltonian $\tilde{H}$ obtained in Step 2 and get $\tilde{H}_\mr{eff}$ defined on qubits.
This introduces some meaningless eigenvalues and even changes the ground-state energy and the low-energy eigenvalues.
For instance, we consider a two-qubit Hamiltonian
\begin{equation}
    \tilde{H}_a = Z \otimes I + I \otimes P^{z+} + P^{x+} \otimes P^{x+},
\end{equation}
where $P^{x +} = (I+X)/2$ [$P^{z +} = (I+Z)/2$] is a projection to an up-spin state in $x$-direction [in $z$-direction].
We add a dimension by one in the same way as Eqs. \eqref{eq: hamiltonian_eff_1}-\eqref{eq: hamiltonian_eff_3}:
\begin{equation}
 \tilde{H}_{a,\mr{eff}} = \tilde{Z} \otimes \tilde{I} + \tilde{I} \otimes \tilde{P}^{z+} + \tilde{P}^{x+} \otimes \tilde{P}^{x+}
\end{equation}
with $\tilde{Z}=Z \oplus (0)$, $\tilde{P}^{\alpha +}=P^{\alpha +} \oplus (0)$, and $\tilde{I}=I \oplus (1)$ (note that the dimension itself is not essential here). While the energy spectrum of $\tilde{H}_a$ is $E \in [-0.836,0.201,1.245,2390]$, that of $\tilde{H}_{a,\mr{eff}}$ is $ [-1,-0.836,0,0,0.201,1,1,1.245,2.390]$, changing the ground state.
As well, another example is $\tilde{H}_b=0.2Z\otimes I + 0.7 I \otimes Z+ 0.3 X \otimes X$.
It originally has the first-excited-state energy $E_1=-0.583$, but that of $\tilde{H}_{b,\mr{eff}}$ is $-0.7$.

These meaningless eigenvalues come from inserting the identity by $\tilde{I} = I \oplus (1)$, which gives meaningless energies to states in the auxiliary dimensions.
We note that inserting the identity by $\tilde{I}^\prime = I \oplus 0_M$ generates no meaningless eigenvalues other than zero.
However, the operator $\tilde{I}^\prime$ is decomposed into a number of terms composed of many Pauli $Z$ operators.
This makes $\tilde{H}_\mr{eff}$ have a lot of non-local Pauli operators among the subsystems, so the extension by $\tilde{I} = I \oplus 0_M$ is not practical for performing the VQE for the effective Hamiltonian in Step 3.
In the original Deep VQE, to avoid the change of the ground-state energy (and the low-energy eigenvalues), the variational quantum circuits should be chosen so that all the components in the additional Hilbert space become zero.

We propose an alternative way to construct the effective qubit model so that $\tilde{H}$ and $\tilde{H}_\mr{eff}$ have the same low-energy spectrum, thereby enabling the use of any variational quantum circuit.
Concretely, we specify the terms in the auxiliary dimensions as follows:
\begin{eqnarray}
\tilde{H}_{\text{eff}} &=& \sum_{i=1}^{N_{\text{sub}}} \tilde{H}_{i,\text{eff}} + \sum_{i \neq j}^{N_{\text{sub}}} \sum_\alpha  \nu_{ij}^\alpha \tilde{V}_{i,\text{eff}}^\alpha \otimes \tilde{W}_{j,\text{eff}}^\alpha, \label{eq: hamiltonian_eff_modified_1} \\
\tilde{H}_{i,\text{eff}} &=& \tilde{H}_{i} \oplus \lambda_i I_{2^{N_{\text{eff}}}-K},
\label{eq: hamiltonian_eff_modified_2} \\ 
\tilde{V}_{i,\text{eff}}^\alpha &=& \tilde{V}_{i}^\alpha \oplus 0_{2^{N_{\text{eff}}}-K}, \quad \tilde{W}_{j,\text{eff}}^\alpha = \tilde{W}_{j}^\alpha \oplus 0_{2^{N_{\text{eff}}}-K}. \label{eq: hamiltonian_eff_modified_3}
\end{eqnarray}
The difference from Eq.~\eqref{eq: hamiltonian_eff_1} is the insertion of $\lambda_i I_{2^{N_{\text{eff}}}-K}$ in $\tilde{H}_{i,\text{eff}}$ with $\lambda_i > 0$, corresponding to a penalty term to components in auxiliary dimensions.  For $\tilde{H}$ and $\tilde{H}_\mr{eff}$ having the same low-energy spectrum, we should choose a proper set $\{ \lambda_i \}_{i=1}^{N_\mr{sub}}$. We have derived the following proposition, which gives a mathematically rigorous sufficient condition on the choice of $\lambda_i$.

\begin{proposition2*}
Let $E_n (H)$ denote the $n$-th smallest eigenvalue of a Hamiltonian $H$ with the dimension $d$ ($n=0,1,2,\hdots, d-1$). When we choose $\{ \lambda_i \}_i$ in Eq. (\ref{eq: hamiltonian_eff_modified_2}) to satisfy
\begin{equation}\label{eq:lambda_bound}
    \lambda_i > e_{\tilde{H}}(i) + E_n (\tilde{H}) - E_0 (\tilde{H}),
\end{equation}
the spectrum of $\tilde{H}$ coincides that of $\tilde{H}_\mr{eff}$, defined by Eqs. (\ref{eq: hamiltonian_eff_modified_1})-(\ref{eq: hamiltonian_eff_modified_3}), up to the $n$-th smallest eigenvalue:
\begin{equation}\label{eq:same_spectrum}
    E_m (\tilde{H}) = E_m (\tilde{H}_\mr{eff}) \quad \mr{for} \quad m=0,1,\hdots, n.
\end{equation}
Here, $e_{\tilde{H}}$ is the value called extensiveness of the $i$-th subsystem under the Hamiltonian $\tilde{H}$, representing the maximal energy of the $i$-th subsystem [see Appendix~\ref{Asec:construction} for the rigorous definition].
When the Hamiltonian $\tilde{H}$ is given by Eq.~\eqref{eq: hamiltoniantilde}, the extensiveness is given by
\begin{equation}
    e_{\tilde{H}}(i) = ||\tilde{H}_i||_\mr{op} + \sum_{j,k: \, \{j,k\} \ni i} \sum_\alpha  |\nu_{jk}^\alpha| \cdot ||\tilde{V}_j^\alpha||_\mr{op} \cdot ||\tilde{W}_k^\alpha||_\mr{op}.
\end{equation}
\end{proposition2*}
We give a proof for this proposition in Appendix~\ref{Asec:construction}.

We discuss the implication of this proposition.
When we calculate the ground state by our modified Deep VQE with Eqs.~\eqref{eq: hamiltonian_eff_modified_1}-\eqref{eq: hamiltonian_eff_modified_3} and choose $\lambda_i$ larger than $e_{\tilde{H}}(i)$, Equation~\eqref{eq:same_spectrum} indicates that we can find the ground state of $\tilde{H}$ by searching that of $\tilde{H}_\mr{eff}$ in the whole Hilbert space of $2^{N_\mr{eff}}$ dimension.
In other words, we can exploit any variational quantum circuits for VQE on the extended Hamiltonian $\tilde{H}_\mr{eff}$.
Similarly, when we calculate the excited states of $\tilde{H}_\mr{eff}$ up to $n$-th level and choose $\lambda_i$ larger than $e_{\tilde{H}}(i)+E_n({\tilde{H}})-E_0(\tilde{H})$, the spectra of $\tilde{H}_\mr{eff}$ and $\tilde{H}$ coincide and we can use an arbitrary variational quantum circuit.

The value of $e_{\tilde{H}}(i)$ can be calculated by classical computers as a sum of $O(N_\mr{sub} \cdot K^3 )$ terms since the minimal or maximal eigenvalue of $K \times K$ matrices is required for evaluating the operator norm.
On the other hand, the energy gap $E_n({\tilde{H}})-E_0(\tilde{H})$ is not known {\it a priori} before performing the VQE for the excited states of $\tilde{H}_\mr{eff}$.
In practice, we can obtain rough and typical energy scale $\Delta_n^\ast \sim E_n({\tilde{H}})-E_0(\tilde{H})$ with other computationally-light methods such as a perturbation theory and a mean-field theory.
With replacing the gap $E_n({\tilde{H}})-E_0(\tilde{H})$  in Eq. \eqref{eq:lambda_bound} by a certain value larger than the estimated gap $\Delta^\ast$, we can safely search the low-energy excited states by VQE on $\tilde{H}_\mr{eff}$ with any variational quantum circuit. 

In short, by constructing the effective qubit model $\tilde{H}_\mr{eff}$ as Eqs.~\eqref{eq: hamiltonian_eff_modified_1}-\eqref{eq: hamiltonian_eff_modified_3} and setting $\lambda_i$ to satisfy Eq.~\eqref{eq:lambda_bound}, we can employ any variational quantum state to run algorithms in the literature~\cite{McClean2017hybrid,nakanishi2018subspace,Parrish2019,higgott2019variational,jones2019variational, Ollitrault2020quantum} to find excited states of $\tilde{H}_\mr{eff}$, which are ensured to be identical to those of $\tilde{H}$. We also note that introducing the energy shift $\lambda_i$ hardly affects the number of measurements. The energy shift terms $0_K \oplus \lambda_i I_{2^{N_{\mr{eff}}}-K}$, composed of $ZZ$-type Pauli operators, have at-most $2^{N_{\mr{eff}}}$ Pauli terms. This is much less than $O(4^{N_{\mr{eff}}})$, required for the effective Hamiltonian $\tilde{H}_i \oplus 0_{2^{N_{\mr{eff}}}-K}$.

\subsection{\label{subsec:protocol} Protocol and number of measurements}

\begin{figure}
\begin{center}
    \includegraphics[height=5cm, width=9cm]{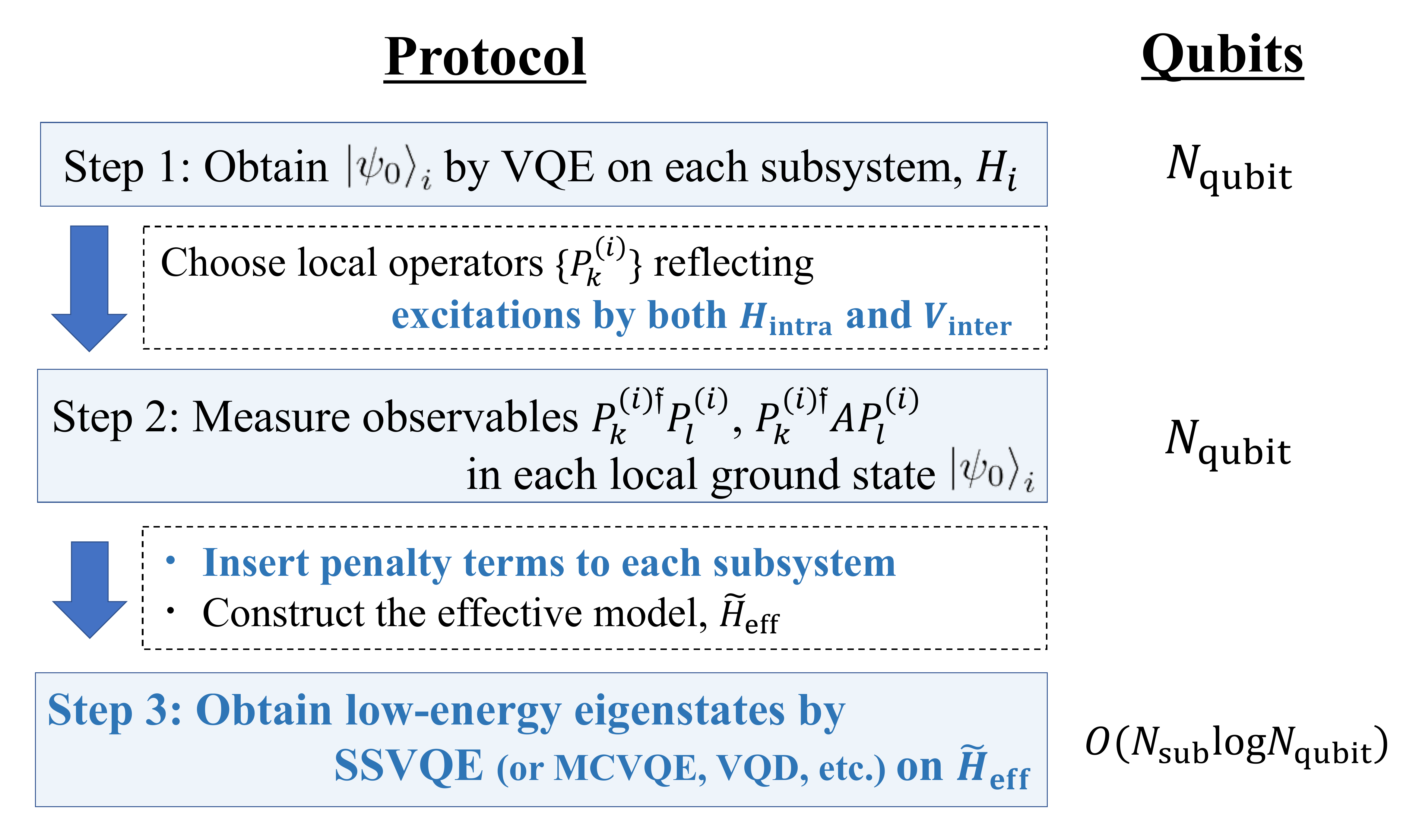}
    \caption{Protocol of the Deep VQE modified for low-lying eigenstates. The blue bold parts are the differences compared to the original Deep VQE \cite{fujii2020deep}, which deals with the ground state. ``Qubits" gives the typical size of NISQ devices required for each step.} 
    \label{fig:protocol}
 \end{center}
 \end{figure}

Here, we summarize the protocol of the modified Deep VQE for low-lying eigenstates and the number of measurements in each step. Figure \ref{fig:protocol} shows a protocol on NISQ devices required for each step.

Let us consider a simple Hamiltonian $H$ involving at-most two-body interactions. In Step 1, the VQEs on subsystems require measurements of $O(N_\mr{sub}N_\mr{qubit}^2)$ terms in $H_\mr{intra}$. In Step 2, we pick up $\{ P_k^{(i)} \}$, which reflects the local excitations from $H_\mr{intra}$ and $V_\mr{inter}$. For a simple choice of $\{ P_k^{(i)} \}$ given by Eq. (\ref{eq:localop_bulk}) with $K \sim N_\mr{qubit}$, which captures linear excitations, we should measure $P_k^{(i)\dagger}P_l^{(i)}$ and $P_k^{(i)\dagger} A P_l^{(i)}$ in $\ket{\psi_0}_i$ [see Eqs. (\ref{eq:overlap}) and (\ref{eq:matrix_elements})]. They involve at-most $O(N_\mr{sub}^2 N_\mr{qubit}^3)$ Pauli terms. In Step 3, we analyze the effective Hamiltonian $\tilde{H}_\mr{eff}$ by elaborated VQE for low-lying eigenstates, such as the subspace-search VQE (SSVQE)~\cite{nakanishi2018subspace}, the multistate-contracted VQE (MCVQE)~\cite{Parrish2019}, and the variational quantum deflation (VQD)~\cite{higgott2019variational}. Since each of $\tilde{H}_i$, $\tilde{V}_i^\alpha$, $\tilde{W}_i^\alpha$ is decomposed into at-most $4^{N_\mr{eff}} \sim K^2$ Pauli terms, we should measure $O(N_\mr{sub}^2 N_\mr{qubit}^4)$ terms in $\tilde{H}_\mr{eff}$. We note that introducing the penalty term $\lambda_i$ by Eq. (\ref{eq: hamiltonian_eff_modified_2}) does not severely increase the cost of measurements. While the number of measurements increases compared to the conventional VQE for the whole system, given by $O(N^2)=O(N_\mr{sub}^2 N_\mr{qubit}^2)$, the size of quantum devices  decreases to $O(N_\mr{sub} \log N_\mr{qubit})$ in the modified Deep VQE.

In the numerical simulations below, we employ SSVQE in Step 3 due to its simplicity, and we give detailed description of its algorithm in Appendix \ref{Aseq:ansatz}. 

\begin{table*}[]
    \centering
    \begin{tabular}{|c|c||c|c|c|c|c|c|c|c|} \hline
       & $\{ P_k^{(i)} \}_k$ & $E_0$(Effective) & $E_0$(Deep VQE) & $E_0$(ED) & $E_1$(Effective) & $E_1$(Deep VQE) & $E_1$(ED) & TR & $N_\mr{req}$ \\ \hline 
       $N_\mr{sub}=2$ & $\mathcal{W}_1$ & $-13.445$ / $0.4$\% & $-13.445$ / $0.4$\% & $-13.500$ & $-11.169$ / $6.4$\% & $-11.169$ / $6.4$\% & $-11.929$ & $6.3 \times 10^{-2}$ & 4 \\ \cline{2-4} \cline{6-7} \cline{9-10}
       $N_\mr{qubit}=4$ & $\mathcal{W}_2$ & $-13.497$ / $0.02$\% & $-13.488$ / $0.09$\% & & $-11.882$ / $0.4$\% & $-11.863$ / $0.6$\% & & $3.9 \times 10^{-1}$ & 8 \\ \hline
       $N_\mr{sub}=3$ & $\mathcal{W}_1$ & $-20.413$ / $0.8$\%  & $-20.413$ / $0.8$\% & $-20.568$ & $-18.665$ / $4.0$\% & $-18.665$ / $4.0$\% & $-19.445$ & $2.7\times 10^{-2}$ & 7  \\ \cline{2-4} \cline{6-7} \cline{9-10}
       $N_\mr{qubit}=4$ & $\mathcal{W}_2$ & $-20.513$ / $0.3$\% & $-20.486$ / $0.4$\% & & $-19.265$ / $0.9$\% & $-19.199$ / $1.3$\% & & $2.4 \times 10^{-1}$ & 12 \\ \hline
       $N_\mr{sub}=2$ & $\mathcal{W}_1$ & $-20.480$ / $0.4$\% & $-20.480$ / $0.4$\% & $-20.568$ & $-18.286$ / $6.0$\% & $-18.286$ / $6.0$\% & $-19.445$ & $3.9\times 10^{-3}$ & 4 \\ \cline{2-4} \cline{6-7} \cline{9-10}
       $N_\mr{qubit}=6$ & $\mathcal{W}_2$ & $-20.560$ / $0.04$\% & $-20.551$ / $0.08$\% & & $-19.343$ / $0.5$\% & $-19.324$ / $0.6$\% & & $6.3 \times 10^{-2}$ & 8 \\ \hline
       $N_\mr{sub}=2$ & $\mathcal{W}_1$ & $-27.535$ / $0.4$\% & $-27.535$ / $0.4$\%  & $-27.647$ & $-25.374$ / $5.2$\% & $-25.374$ / $5.2$\% & $-26.770$ & $2.4\times 10^{-4}$ & 4 \\ \cline{2-4} \cline{6-7} \cline{9-10}
       $N_\mr{qubit}=8$ & $\mathcal{W}_2$ & $-27.634$ / $0.05$\% & $-27.612$ / $0.1$\% & & $-26.620$ / $0.6$\% & $-26.578$ / $0.7$\% & & $7.4 \times 10^{-3}$ & 10 \\ \hline
    \end{tabular}
    \caption{Numerical results for low-energy eigenvalues $E_n$ ($n=0,1$) of a one-dimensional AFM Heisenberg model~\eqref{eq: AFM model}.
    The excitation operator set $\{P_k^{(i)}\}$ is taken as $\mathcal{W}_1$ (by $\mathcal{W}_2$) for the original Deep VQE (for our modified Deep VQE).
    The values $E_n (\mr{Deep VQE})$ and $E_n (\mr{ED})$ represent the results of Deep VQE simulation and those of exact diagonalization of the original Hamiltonian $H$.
    The values $E_n (\mr{Effective})$ are obtained by substituting ED for the VQE in Step 3 of Deep VQE, which give the best performance of Deep VQE in theory.
    The left values and the right values of $E_n(\mr{Effective})$ and $E_n(\mr{Deep VQE})$ are the obtained energies and their relative errors from $E_n(\mr{ED})$, respectively.
    The truncation rate TR is given by Eq. (\ref{eq:TR}). We also show the number of qubits required to run Deep VQE by $N_\mr{req}$. }
    \label{tab:spin_chain}
\end{table*}

\subsection{\label{subsec:spinchain} Example: Spin chain}
Based on the modified Deep VQE protocol in the previous subsections, we numerically examine its validity by exemplifying a simple spin model.
We consider a one-dimensional anti-ferromagnetic (AFM) Heisenberg model of $N_\mr{tot}$-site,
\begin{equation} \label{eq: AFM model}
H = \sum_{i=1}^{N_\mr{tot}-1} (X_i X_{i+1} + Y_i Y_{i+1} + Z_i Z_{i+1}),
\end{equation}
under the open boundary condition.
We split the system into $N_\mr{sub}$ subsystems, each of which is composed of neighboring $N_\mr{qubit}$ qubits ($N_\mr{tot}=N_\mr{sub} \times N_\mr{qubit}$). Then, the inter-subsystem interactions involve qubits only at the boundaries of the subsystems.

We introduce two different sets of local excitation operators to compare the modified Deep VQE with the original Deep VQE. The first one is
\begin{equation}
    \mathcal{W}_1 = \{ I \} \cup \{ X_i, Y_i, Z_i | \text{ $i$ at the boundary of subsystem} \}.
\end{equation}
We note that we have not written the index of the subsystem, but we define $\mathcal{W}_1$ for each subsystem.
An excitation operator $P_k \in \mathcal{W}_1$ only acts on the boundaries of subsystems and is relevant to the inter-subsystem interactions, so the choice of local basis based on $\mathcal{W}_1$ correspond to the original Deep VQE.
The dimension of the local Hilbert space, $K$, is $4$ (for subsystems at the edges) or $7$ (for subsystems in the bulk).
The second choice of local excitation operators is
\begin{equation}
    \mathcal{W}_2 = \{ I \} \cup \{ X_i, Y_i, Z_i | \,i \in \Lambda^\prime \},
\end{equation}
where $\Lambda^\prime$ is a set of qubits in a subsystem except for the right edge. This choice also captures any single-spin excitation $(\alpha X_i + \beta Y_i + \gamma Z_i) \ket{\psi_0}$ ($\alpha,\beta,\gamma \in \mathbb{C}$), emerging from a generic local extensive intra-subsystem Hamiltonian $H_i$ as well as those of $V_\mr{inter}$, thereby corresponding to the choice in our modified Deep VQE.
We omit qubits at the right edge because of the SU(2) symmetry in the AFM model; for the ground state $\ket{\psi_0}$ of each subsystems, it holds $\left(\sum_{j \in \text{subsystem}} X_j \right) \ket{\psi_0} = \left(\sum_{j \in \text{subsystem}} Y_j \right) \ket{\psi_0} = \left(\sum_{j \in \text{subsystem}} Z_j \right) \ket{\psi_0} = 0$.
One of the states in $\{ X_j \ket{\psi_0}| \text{ $j$ in subsystem} \}$ is not linearly independent from the others, and the same relations for $Y$ and $Z$ also hold.
We do not include the Pauli operators ($X,Y,Z$) at the right edge in $\mathcal{W}_2$ due to this fact.
The local Hilbert space dimension $K$ is $3 N_\mr{qubit}-2$ in the case of $\mathcal{W}_2$.

We note on the insertion of auxiliary dimensions in this model analyzed in Sec.~\ref{subsec:modified2}.
We can roughly estimate the extensiveness $e_{\tilde{H}}(i)$ for each subsystem.
In the case of $N_\mr{sub}=2$ and $N_\mr{qubit}=4$, 
we numerically confirm that $||\tilde{H}_i||_\mr{op}$ coincides with the absolute value of the ground-state energy for each subsystem.
The operator norm of the inter-subsystem interactions is at most $2 (||X_i||_\mr{op} ||X_{i+1}||_\mr{op}+||Y_i||_\mr{op} ||Y_{i+1}||_\mr{op}+||Z_i||_\mr{op} ||Z_{i+1}||_\mr{op})=6$, so we obtain $e_{\tilde{H}}(i) \leq 6.464 + 6 =12.464$.
Therefore, it is sufficient to insert auxiliary dimensions by $J_i=\lambda_i I_{2^{N_\mr{eff}}-K}$ with $\lambda_i > 12.464$ (for ground states) or $\lambda_i > 12.464 + \Delta_n^\ast$ (for excited states) with the estimated gap $\Delta_n^\ast \sim E_n(\tilde{H})-E_0(\tilde{H})$.
Nevertheless, we directly confirm that both the ground-state energy and the first-excited-state energy of $\tilde{H}$ are the same as those of $\tilde{H}_\mr{eff}$ by exact diagonalization in this simulation, so we simply set $\lambda_i=0$ in the following.

We run a classical simulation of Deep VQE for the AFM Heisenberg model~\eqref{eq: AFM model}.
We exploit a hardware-efficient type ansatz~\cite{kandala2017hardware,Mitarai2018quantum} as a variational quantum circuit.
We use this ansatz in the VQE for finding the ground states of the subsystem Hamiltonians (Step 1 of Deep VQE) and SSVQE for finding the ground and first-excited state of the effective Hamiltonian $\tilde{H}_\mr{eff}$ (Step 3 in Deep VQE).
The simulation of quantum circuits is performed by using the libraries Qulacs~\cite{qulacs_2018, suzuki2020qulacs} and OpenFermion~\cite{mcclean2017openfermion}.
More details on numerical simulations are described in Appendix~\ref{Aseq:ansatz}.

We show the numerical results for the AFM Heisenberg chain with various partitioning of the total system in Table \ref{tab:spin_chain}.
We calculate the ground-state energy $E_0$ and the first-excited-state energy $E_1$ in three ways and compare them.
The first one, ``Effective" is obtained by exact diagonalization of the effective Hamiltonian $\tilde{H}$.
This value tells us provides the best performance of Deep VQE in theory.
The second one, ``Deep VQE," is obtained by simulations of Deep VQE.
The difference between ``Effective" and ``Deep VQE" indicates errors originating from VQEs in Step 1 and 3 of Deep VQE.
The last one, ``ED," is the exact energy of of the original Hamiltonian $H$ calculated by the exact diagonalization.
The difference between ``Effective" and ``ED" tells the validity of the choice of the local basis in Step 2 of Deep VQE.
In the table, we also introduce the truncation rate $\mr{TR}$ defined by
\begin{equation}\label{eq:TR}
    \mr{TR} = \mr{dim}(\tilde{\mathcal{H}})/\mr{dim}(\mathcal{H}) = \left( K/ 2^{N_\mr{qubit}} \right)^{N_\mr{sub}},
\end{equation}
as a figure of merit of the reduction of the Hilbert space.
The accurate results for $E_n(\text{Effective})$ ($n=0,1$) with small $\mr{TR}$ mean that the coarse-graining is performed with a proper choice of a local basis set. 

As shown in Table~\ref{tab:spin_chain}, in terms of the ground-state energy $E_0$, both the original Deep VQE ($\mathcal{W}_1$) and the modified Deep VQE ($\mathcal{W}_2$) exhibit good performance with relative errors to ``ED" values (eigenvalues of the original Hamiltonian) up to $0.8$\%, as expected.
On the other hand, in terms of the first-excited-state energy $E_1$, the modified Deep VQE ($\mathcal{W}_2$) shows comparably good performance with relative errors from $0.6 \%$ to $1.3 \%$, compared to the original Deep VQE ($\mathcal{W}_1$) results having relative errors from $4.0 \%$ to $6.4 \%$.
Considering that the value of ``Effective" in the original Deep VQE ($\mathcal{W}_1$) deviates from that of ``ED", the modified choice of local basis is essential for obtaining low-energy excited states accurately, i.e., the larger error in the original Deep VQE ($\mathcal{W}_1$) is not due to the imperfection of the optimization of the VQE in Deep VQE algorithm. We also evaluate the number of measurements. For $N_\mr{sub}=2$ and $N_\mr{qubit}=8$ with $\mathcal{W}_2$, each subsystem term $\tilde{H}_i \oplus 0_{2^{N_{\mr{eff}}}-K}$ is decomposed into $452$ Pauli terms. While we do not introduce the energy shift $\lambda_i$ for the simulation, we can complete the protocol with the same number of measurements $452$ for $\tilde{H}_i \oplus \lambda_i I_{2^{N_{\mr{eff}}}-K}$.

To further understand the higher accuracy for the excited states achieved with our local basis choice $\mathcal{W}_2$, we confirm the validity of QSE for each subsystem as discussed in Section \ref{subsec:modified1}. QSE for each subsystem with the local excitations $\mathcal{W}_1$ ($\mathcal{W}_2$) and the reference state $\ket{\psi_0}_i$ is equivalent to calculating the eigenvalues of $\tilde{H}_i$ obtained from $\mathcal{W}_1$ ($\mathcal{W}_2$) and $\ket{\psi_0}_i$.
In the case of  $N_\mr{qubit}=6$, the exact first-excite-state energy for each subsystem is $E_1^\mr{sub}=-8.008$ with three-fold degeneracy, which we calculate by the original $H_i$. On the other hand, by diagonalizing $\tilde{H}_i$, we obtain the QSE results $E_1^\mr{sub}=-6.415$ with nearly three-fold degeneracy (for $\mathcal{W}_1$) and  $E_1^\mr{sub}=-8.000$ with three-fold degeneracy (for $\mathcal{W}_2$). This means the failure of QSE with $\mathcal{W}_1$ and the success of QSE with $\mathcal{W}_2$.

As discussed in Section \ref{subsec:modified1}, the success of QSE in expressing the local first-excited states ensures the accurate first-excited-state energy by the consistence with the perturbation theory. The above result on the invalidity and validity of QSE for $\mathcal{W}_1$ and $\mathcal{W}_2$ explains why our modified local excitations $\mathcal{W}_2$ gives much better approximation for the excited states than the original one $\mathcal{W}_1$. This example also supports the scenario of the modified Deep VQE in Section \ref{subsec:modified1}---the modified local excitations reflecting the intra-subsystem Hamiltonians result in the accurate low-lying excited states due to the success of QSE for each subsystem. Based on this finding, although we simply use all the Pauli operators $X,Y,Z$, we can find a better choice which efficiently describes each subsystem by employing some methods reflecting the type of intra-subsystem Hamiltonians $H_i$ such as Ref. \cite{Bharti2020Iterative}.



\subsection{\label{subsec:discuss_acc} Discussion for improving the accuracy}
We finally discuss how to improve the accuracy of the low-energy eigenvalues obtained by our modified Deep VQE.
The key is to capture higher-order excitations caused by intra-subsystem and inter-subsystem interactions within each subsystem.
For example, the product of Pauli operators like $A_i B_j$ ($A, B = X, Y, Z$) for different qubits $i,j$ within the same subsystem can be included in the local operator set in Step 2 of Deep VQE, while our simulation only considers linear excitations of such Pauli operators.
This is similar to considering higher-order excitations in QSE.
The drawback for including the higher-order excitaions is the increase of the effective system size.
When we consider up to the $n$-th order product of Pauli operators like $P_{j_1}\cdots P_{j_n}$, the dimension of the local Hilbert becomes $K \sim (N_\mr{qubit})^n$.
Although we have $(N_\mr{qubit})^n \ll 2^{N_\mr{qubit}}$ (the dimension of the original local Hilbert space) for small $n$, the order of excitations $n$ has to be determined by comparing the benefit of improving the accuracy and the cost of the computation.


\section{\label{sec:chemistry} Main Result: Application to chemistry problems}

\begin{figure}
\begin{center}
    \includegraphics[height=4.5cm, width=9cm]{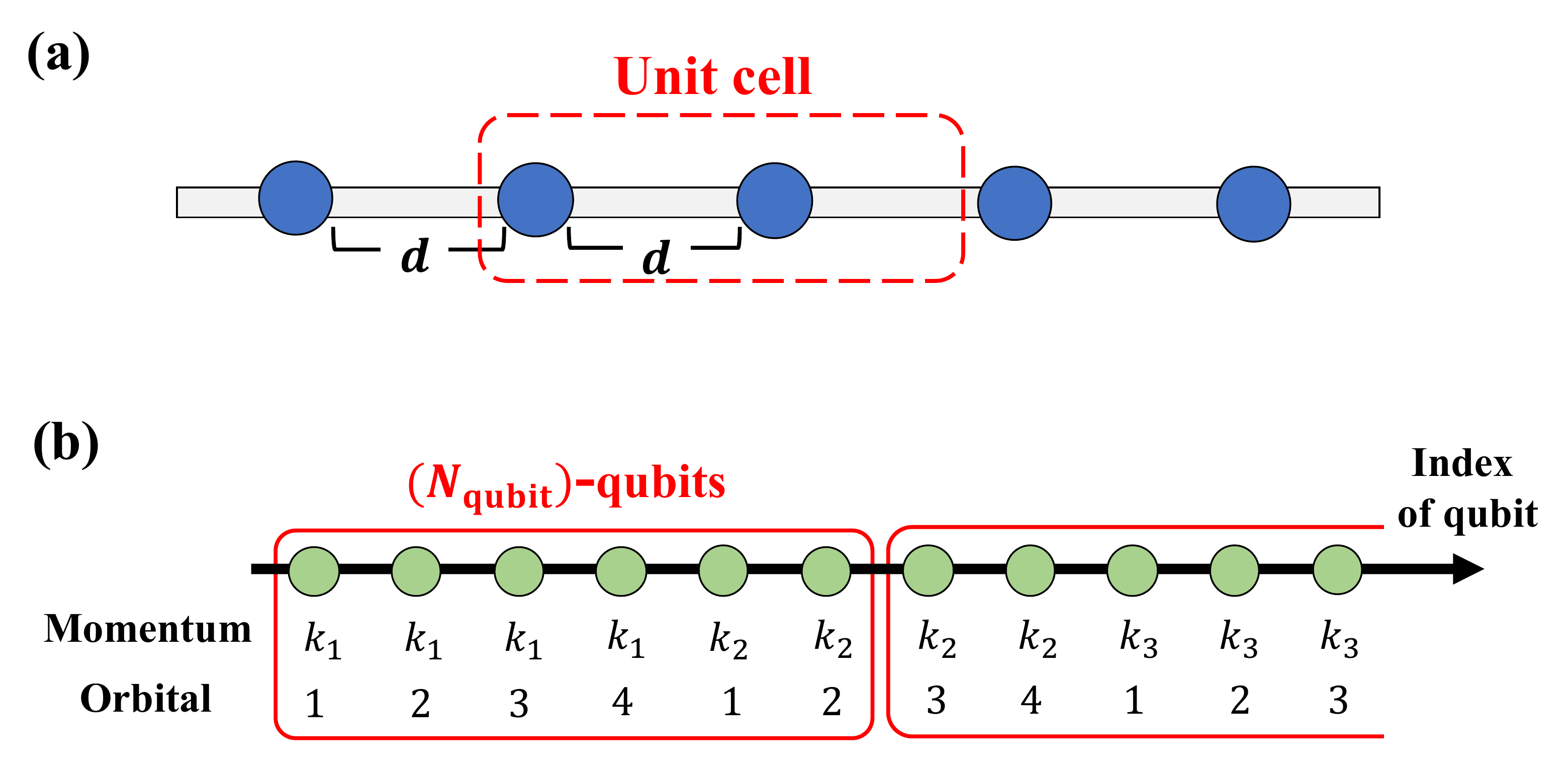}
    \caption{(a) A one-dimensional hydrogen chain. A unit cell includes two hydrogen atoms, and the distance between the atoms is $d$. (b) The way of splitting the hydrogen chain into subsystems in the momentum space.
    Neighboring $N_\mr{qubit}$ qubits, corresponding to electrons with similar momentum, belong to the same subsystem. Here, the orbitals $1$ and $2$ ($3$ and $4$) represent the occupied (unoccupied) crystalline orbitals under the Hartree-Fock approximation.} 
    \label{fig:hydrochain}
 \end{center}
 \end{figure}

In this section, we provide the second half of our main results.
We apply our modified VQE to the simplest example of a quantum chemistry calculation for periodic material: electronic states of a periodic hydrogen chain.
We show numerical simulation results and  examine the validity of our modified Deep VQE on the low-energy eigenstates of periodic materials.
Our numerical results also can be seen as the first application of Deep VQE to fermionic systems that map to non-local qubit Hamiltonians in general.


\subsection{\label{subsec:model_method} Model and Method}
Let us describe our model and how to perform Deep VQE. We consider a one-dimensional chain under periodic boundary condition whose unit cell is composed of two hydrogen atoms [see Fig. \ref{fig:hydrochain} (a)].
The distance between the two atoms is $d$ and the length of the unit cell is $2d$.
After performing the crystal
Hartree–Fock calculation with STO-3G basis set~\cite{yoshioka2020periodic}, we obtain 
the Hamiltonian in the second-quantized form as
\begin{eqnarray}
    H &=& \sum_k \sum_{pq} t_k^{pq} c_{kp}^\dagger c_{kq} \nonumber \\
    &\quad& + \sum_{k_1 k_2 k_3 k_4} {}^\prime \sum_{pqrs} v_{k_1 k_2 k_3 k_4}^{pqrs} c_{k_1 p}^\dagger c_{k_2 q}^\dagger c_{k_3 r} c_{k_4 s}, \label{eq:hamiltonian_hydrogen}
\end{eqnarray}
where $c_{kp}^\dagger$ ($c_{kp}$) is a creation annihilation operators of electrons with crystalline momentum $k$ and the spin orbital $p$.
The crystalline momentum $k$ is uniformly sampled from the first Brillouine zone $[0,2\pi]$, which is renormalized by the unit-cell length $2d$.
The index $p \in [1,2,3,4]$ represents the spin-orbital within a unit cell, which comes from two spin-orbitals in the STO-3G basis for two hydrogen atoms.
The orbitals $p=1,2$ ($p=3,4$) are occupied (unoccupied) orbitals in the Hartree-Fock state.
The symbol $\Sigma^\prime$ represents the summation over the crystalline momentum under the crystalline momentum conservation,  satisfying
\begin{equation}\label{eq:momentum_conserve}
    k_1 + k_2 - k_3 - k_4 \in 2 \pi \mathbb{Z}.
\end{equation}
The sets of coefficients $\{ t_k^{pq} \}$ and $\{ v_{k_1 k_2 k_3 k_4}^{pqrs} \}$ represent one-body and two-body electron integrals between different crystalline Hartree-Fock orbitals $(k,p)$, determined by classical computers.

To perform Deep VQE, we transform the fermionic Hamiltonian~\eqref{eq:hamiltonian_hydrogen} into the one on qubits by the Jordan-Wigner transformation \cite{Jordan1928}.
Each qubit is still labeled by $(k, p)$ of a corresponding electron.
Through this transformation, some non-local terms appear in the Hamiltonian, e.g., a Pauli string $X_1 Z_2 X_3 X_5 Z_6 X_7$ coming from $c_3^\dagger c_7^\dagger c_1 c_5$.
This non-locality potentially harms the validity of Deep VQE because the eigenstates can be entangled non-locally among the system, so our simulation also examines how the non-locality affects the accuracy of Deep VQE.
When we sample $N_k$ points as the crystalline momentum, $4 N_k$ qubits are required to represent the original Hamiltonian~\eqref{eq:hamiltonian_hydrogen}.

Next, we depict how to perform each step of Deep VQE for the periodic hydrogen chain including the way to divide it into subsystems.
We define subsystems so that each orbital (qubit) in the subsystem has similar the crystalline momentum $k$.
To be precise, we label the qubits as $(k,p)=(k_1,1),(k_1,2),(k_1,3),(k_1,4),(k_2,1),(k_2,2),\hdots$ with the order $0=k_1<k_2<\hdots$ [see Fig. \ref{fig:hydrochain} (b)] and group neighboring $N_\mr{qubit}$ qubits with the smaller index step by step.
We stress that we define the subsystems in the momentum space, not the real space.

We consider two set of local excitation operators for Step 2 of Deep VQE as
\begin{eqnarray}
    \mathcal{W}_\mr{s} &=& \{I\} \cup \{ c_j^\prime, \, c_j^{\prime \dagger} \, | \, j \in \Lambda_e \}, \label{eq:single_excitation} \\
    \mathcal{W}_\mr{d} &=& \mathcal{W}_\mr{s} \cup \{ c_j^{\prime \dagger} c_k^\prime \, | \, j,k \in \Lambda_e \}, \label{eq:double_excitation}
\end{eqnarray}
where $\Lambda_e$ is the set of $(k,p)$ in each subsystem (we have not explicitly written the index for the subsystem in $\mathcal{W}_s$ and $\mathcal{W}_d$).
The qubit operator $c_i^\prime$ ($c_i^{\prime \dagger}$) is obtained by a fermion annihilation (creation) operator truncated within each subsystem.
Namely, we define $c_i^\prime$ ($c_i^{\prime \dagger})$ by restricting the transformed Pauli operators of $c_i$ $(c_i^\dag)$ to act only on the target subsystem.
For instance, in the case of $N_\mr{qubit}=4$, we define $c_6^\prime = Z_5 (X_6 - i Y_6)/2$ while the genuine qubit-operator representation of $c_6$ obtained by the Jordan-Wigner transformation is $c_6=Z_1 Z_2 Z_3 Z_4 Z_5 (X_6-iY_6)/2$.
Although this truncation means neglecting the anti-commutation relations between $c_i^\prime$ and $c_i^{\prime \dagger}$ in different subsystems, we expect that they are sufficient to describe locally excited states in each subsystem.
Under this definition, $\mathcal{W}_\mr{s}$ is a set of single-particle excitations giving a local Hilbert space dimension $K=|\mathcal{W}_\mr{s}|=2N_\mr{qubit}+1$.
On the other hands, $\mathcal{W}_\mr{d}$ is a set of single- and double- particle excitations giving a local Hilbert space dimension  $K=(N_\mr{qubit}+1)^2$.

In Step 3 of Deep VQE, we should specify the way of inserting auxiliary dimensions following Eqs. (\ref{eq: hamiltonian_eff_modified_1})-(\ref{eq: hamiltonian_eff_modified_3}).
We again employ $\lambda_i=0$ because we directly confirm that the low-energy eigenvalues of $\tilde{H}_\mr{eff}$ and those of $\tilde{H}$ coincide by the exact diagonalization.

\subsection{\label{subsec:hydrogen_results} Numerical results}

\begin{table}[]
    \centering
    \begin{tabular}{|c|c|c|c|} \hline
    $E_0$(Local) & $E_0$(Effective: ED) & $E_0$(Deep VQE) & $E_0$(ED) \\ \hline
     $-0.743$ / $31$\%  & $-1.067$ / $1.4$\% & $-1.067$ / $1.4$\% & $-1.082$ \\ \hline \hline
    --- & $E_1$(Effective: ED) & $E_1$(Deep VQE) & $E_1$(ED) \\ \hline
    --- & $-0.743$ / $4.1$\% & $-0.743$ / $4.1$\% & $-0.775$ \\ \hline
    \end{tabular}
    \caption{Numerical results for the ground-state and the first-excited-state energies of the periodic hydrogen chain under $d=1.4$ Bohr. The left values in the cells represent energies delivered in Hartree. The right values are relative errors from the exact results $E_n$(ED).}
    \label{tab:hydro_chain}
\end{table}

\begin{figure*}
\begin{center}
    \includegraphics[height=8.5cm, width=18cm]{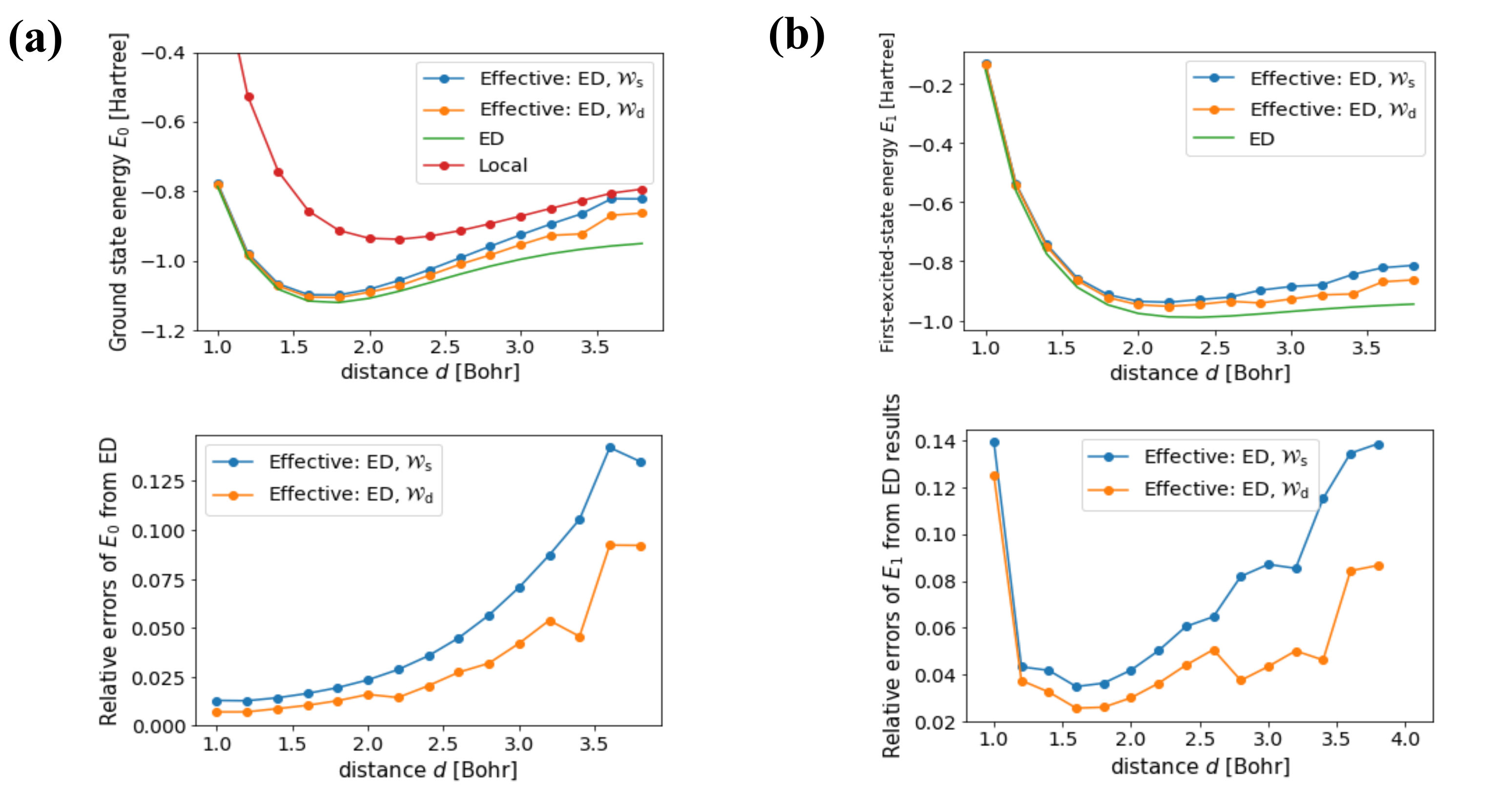}
    \caption{Numerical results for the hydrogen chain with picking up three $k$-points. The values of ``ED'' and ``Local'' show the exact and first-order-perturbation~\eqref{eq: E0_local} results, respectively.
    ``Effective: ED'' results are obtained by substituting VQEs with ED in Steps 1 and 3 of Deep VQE, giving the best performance of Deep VQE in theory. (a) The ground-state energy [upper panel] and the relative errors from the ED values [lower panel]. (b) The first-excited-state energy [upper panel] and the errors from the ED values [lower panel].
    For both the ground- and first-excited-state energies, the relative errors from the exact results with double-particle excitations are suppressed approximately half compared to those with single-particle excitations.  } 
    \label{fig:hydrograph}
 \end{center}
 \end{figure*}
 
In numerical simulations of Deep VQE for the periodic hydrogen chain, three $k$-points are sampled (the total number of qubits is $3\times4=12$), and we split the system by $N_\mr{sub}=2$ and $N_\mr{qubit}=6$.
Here, we employ a hardware-efficient type ansatz~\cite{kandala2017hardware,Mitarai2018quantum} of depth $20$ for VQE in Step 1 of Deep VQE, and that of depth $80$ for SSVQE in Step 3 to simply examine the performance of our protocol. To compute larger systems with smaller depth of quantum circuits, it will be better to employ VQD or MCVQE instead of SSVQE (see Fig. \ref{fig:protocol}) and other ansatz avoiding the barren plateau problem \cite{mcclean2018barren} instead of hardware-efficient ansatz.
All circuit simulations are performed by using the libraries Qulacs~\cite{qulacs_2018,suzuki2020qulacs} and OpenFermion~\cite{mcclean2017openfermion}, and the crystalline Hartree-Fock calculation is done by PySCF package~\cite{Sun2018_pyscf,Sun2020} (see Appendix \ref{Aseq:ansatz} for the detail).

To assess the accuracy of Deep VQE result, we introduce two kinds of values, dubbed ``Local" and ``Effective: ED." The former one is defined by
 \begin{equation} \label{eq: E0_local}
     E_0(\mr{local}) = \braket{\Psi_0 | H | \Psi_0 },
 \end{equation}
 where $\ket{\Psi_0}=\bigotimes_{i=1}^{N_\mr{sub}} \ket{\psi_0}_i$ is the ground states of the intra-subsystem Hamiltonian $H_\mr{intra}= \sum_{i=1}^{N_\mr{sub}} H_i$.
 The energy $E_0 (\mr{local})$ represents the result of the first-order perturbation theory with the unperturbed Hamiltonian $H_\mr{intra}$ and the perturbation $V_\mr{inter}=H-H_\mr{intra}$.
 The latter one ``Effective: ED" is calculated by substituting VQE in Step 1 of Deep VQE by exact diagonalization and performing the exact diagonalization again to solve $\tilde{H}_\mr{eff}$ in Step 3 of Deep VQE.
 The value of ``Effective: ED" gives the best performance of Deep VQE in theory since we solve all the Hamiltonians in Deep VQE exactly (Note that its definition is different from that of ``Effective" in Section III in that the latter exploits VQE in Step 1 and ED in Step 3).
 
First, we choose the single-particle excitations $\mathcal{W}_\mr{s}$ as a local basis of Step 2 of Deep VQE and set $d= 1.4$ Bohr.
The results are shown in Table~\ref{tab:hydro_chain}.
We find a nice agreement between the ``Deep VQE" and the ``ED" within relative error up to a few percent, especially compared to the value of the perturbation theory ``$E_0(\mr{local})$".
It should be noted that this accuracy characterized by the relative errors up to $4.1 \%$ is achieved despite truncating the vast majority of the whole Hilbert space, $\mr{TR}=0.0413$, and reducing the number of qubits by four ($12\to8$).
We also find that the values of ``Effective: ED" and those of ``Deep VQE" are almost identical, which implies that the error of the ``Deep VQE" results from the exact ones solely comes from the truncation of the Hilbert space in Step 2 of Deep VQE.

 Next, let us investigate how the validity of Deep VQE depends on the atom-atom distance $d$ and the local basis.
 From now on, we consider the values of $E_n (\text{Effective: ED})$ ($n=0,1$) instead of the Deep VQE results $E_n(\text{Deep VQE})$ due to the computational cost of classical simulations.
 As discussed in the above, the values of $E_n(\mr{Effective: ED})$ give the best performance of the Deep VQE protocol.
 In some cases, we observe the two-fold degeneracy for the ground states of each subsystem, and we randomly choose the local ground states $\ket{\psi_0}_i$ in such situations.
 Figure \ref{fig:hydrograph} shows the low-energy eigenvalues [upper panels] and the relative error [lower panels] of $E_n(\text{Effectice: ED})$ from the exact results $E_n(\text{ED})$ for (a) the ground state and (b) the first-excited states.
 For the ground-state energy, Fig. \ref{fig:hydrograph} (a) shows that the Deep VQE results with the local operators $\mathcal{W}_\mr{s}$ and $\mathcal{W}_\mr{d}$ both can provide much more accurate values than those of the first-order perturbation theory, $E_0(\mr{local})$, within a wide range of the distance $d$.
 When we only consider the single-particle excitations by $\mathcal{W}_\mr{s}$, the obtained ground-state energy is accurate under small distance $d$ while the result approaches $E_0 (\mr{local})$ with increasing $d$.
 By taking into account double-particle excitations with $\mathcal{W}_\mr{d}$, we can improve the accuracy, especially in the large-$d$ regime with making the relative error approximately half.
 The results for the first-excited states have a similar tendency.
 We can see that the first-excited-state energy error is suppressed up to $10 \%$. The improvement of the accuracy by considering double-particle excitations becomes larger as the distance $d$ increases, with giving the errors approximately half as large as those with single-particle excitations.
 
 We remark on the required number of qubits for simulation. For both cases with the single- and double-particle excitations, we observe some states in  $\{ P_k^{(i)} \ket{\psi_0}_i \}_{k=1}^K$ are not linearly independent.
 For the former case $\mathcal{W}_\mr{s}$, the numerically-obtained dimension $K$ of the restricted Hilbert space ranges from $11$ to $13$, giving the Hilbert space truncation rate $\mr{TR}$ [Eq. (\ref{eq:TR})] ranging from $0.030$ to $0.035$ (Note that $\mr{TR}$ obtained here is different from that of the Deep VQE, $0.0413$, due to the degenerate ground states in the local Hilbert space).
 The Deep VQE calculation for the original model [Eq. (\ref{eq:hamiltonian_hydrogen})] of $12$ qubits can be executed with a $8$-qubit quantum device in this case.
 On the other hand, for the latter case $\mathcal{W}_\mr{d}$, we find the numerically-obtained dimension $K$ of the restricted Hilbert space and the truncation rate is $24 \leq K \leq 34$ and $0.15 \leq \mr{TR} \leq 0.22$, respectively.
 The number of the required qubits for the Deep VQE protocol is $11$ for the atom-atom distance $d=3.4$ Bohr and $10$ for the other choices of $d$.

Finally, we discuss two possible directions to improve the accuracy.
The first way is to increase the number of the reference states for the local basis, e.g., taking the local basis like $\{ P_k^{(i)} \ket{\psi_0}_i,\} \cup \{P_k^{(i)} \ket{\psi_1}_i\} \cup \cdots$.
It is reasonable to include low-energy eigenstates of the subsystems other than the ground state for constructing the local basis to capture the low-energy physics in the original system.
We discuss such a protocol in Appendix \ref{Aseq:multi_state}.
The second possible way is to consider higher-order excitations for the construction of local basis set.
We can show that the Deep VQE with higher-order local excitations gives the better result than the combination of QSE and Deep VQE for the whole system 
with lower-order excitations (see Appendix~\ref{Asec: DeepVQE and QSE} for the detail).
As well as Sec.~\ref{subsec:discuss_acc}, the number of qubits for simulation is given by
\begin{equation}
    N_\mr{req} \sim \min (N_\mr{qubit}, n N_\mr{sub} \lceil \log_2 N_\mr{qubit} \rceil ) \ll N_\mr{qubit} \times N_\mr{sub}
\end{equation}
when we consider up to the $n$-th order excitation.
Therefore, as the excitations up to double-particle $\mathcal{W}_\mr{d}$ overwhelms those up to single-particle $\mathcal{W}_\mr{s}$ in terms of accuracy, the local basis with higher order excitations will give more accurate low-lying excited states with keeping the reduction of qubits. 

\section{\label{sec:conclusions} Discussion and Conclusions}

In this paper, we have proposed the improved way for performing Deep VQE, in which we can properly obtain low-energy eigenstates with arbitrary preferable variational quantum circuits on a smaller number of qubits.
We have composed a set of local excitation operators used for the coarse-graining in Deep VQE by focusing on the excitations caused by intra-subsystem interactions.
The perturbation theory and QSE ensure the validity for low-energy eigenstates obtained by such Deep VQE protocols.
We have also provided an alternative way of constructing the effective Hamiltonian defined on qubits, in which we introduce penalty terms to the auxiliary dimensions.
We have derived a rigorous bound on the penalty that makes arbitrary variational quantum states available.
After reformulating Deep VQE for excited states, we have applied it to periodic materials, namely, a periodic hydrogen chain.
We have shown that the low-energy eigenvalues are well reproduced with relative errors up to $O(1) \%$ in a small atom-atom distance regime by splitting the system into subsystems based on crystalline momentum and considering the local basis introduced by single-particle and double-particle excitations.
Our results enlarge the possibility of simulating large systems by a small-sized quantum computers even for excited states of quantum systems including, quantum chemistry.

We provide some comments on how our method outperforms conventional classical methods. Our method searches low-energy eigenstates from the restricted Hilbert space spanned by
\begin{equation}\label{eq:search_space}
\bigotimes_{i=1}^{N_{\mr{sub}}} \left( P_{k_i}^{(i)} \ket{\psi_0}_i \right), \quad k_i = 1,2, \hdots, K.
\end{equation}
Due to the coarse-graining, this can accurately capture strong intra-subsystem correlations and weak inter-subsystem correlations. In Eq. (\ref{eq:search_space}), each pair of subsystems is equivalent to one another. Thus, our method can deal with 1D systems with long-range interactions, which is usually difficult for classical methods based on matrix product states (MPS) \cite{Vidal2003Efficient,Vidal2004Efficient,Schollwock2005dmrg}. This will benefit chemistry problems where the Coulomb interactions play a central role. In addition, the applicability of long-ranged models indicates the validity of our method for higher dimensional systems unlike MPS-based methods. Our method will overwhelm the cluster mean-field theory (MFT), which is a classical method for high dimensional systems combined with the coarse-graining, since it can deal with entanglement between subsystems.

We leave some future directions for this study. First, while our results provide approximate ground-state and first-excited-state energies for periodic materials with relative errors up to a few percents, the ultimate goal is to predict them with the chemical accuracy ($1.6\times 10^{-3}$ Hartree) by Deep VQE.
As discussed in Secs.~\ref{subsec:discuss_acc} and \ref{subsec:hydrogen_results}, the way to improve the accuracy would be to take higher-order excitations into account for the local basis.
Although the models examined in this paper are too small to get benefit from considering higher-order excitations because those excitations exhaust the original (unrestricted) Hilbert space and there is no decrease in the number of qubits, large systems targeted by the NISQ devices with hundreds or thousands of qubits can gain the improvement of the accuracy and the decrease of the number of qubits by considering higher-order excitations.
Second, it is intriguing to study the performance of our modified VQE when we repeat the coarse-graining of Deep VQE many times, as proposed in Ref.~\cite{fujii2020deep}.
The Deep VQE discussed in this paper performs the coarse-graining  once in Step 2.
As the number of repetition of the coarse-graining increases, the Deep VQE can simulate much larger systems, but instead, its effective Hamiltonian gradually becomes non-local and discards the information of the original systems.
It should be an important problem whether the modified Deep VQE with repeating the coarse-graining many times well reproduces low-energy eigenstates of huge systems.
Third, from the practical point of view, the feasibility of our modified VQE in the real NISQ hardware can be further investigated.
For example, analyzing the effect of noise in the NISQ devices on the coarse-graining of Deep VQE is helpful to find how much we alleviate the noise by error mitigation techniques~\cite{temme2017error,endo2018practical,mcardle2019error}.
Another possible obstacle is the problem so-called barren plateau~\cite{mcclean2018barren}, in which the optimization of variational quantum circuits gets difficult for large and deep quantum circuits.
Our results in this paper nevertheless provides the solid support that our modified VQE works in noiseless situations.

\section*{Acknowledgment}
We thank W. Mizukami and K. Fujii for fruitful discussion on VQE for periodic materials and Deep VQE. K. M. is supported by WISE Program, MEXT, and a Research Fellowship for Young Scientists from JSPS (Grants Np. JP20J12930).


%

\appendix
\begin{center}
\bf{\large Appendix}
\end{center}

\section{Multi-state Deep VQE for excited states}\label{Aseq:multi_state}
In this section, we discuss another choice of the local basis in Step 2 of Deep VQE for calculating low-energy excited states.
As discussed in Sec.~\ref{subsec:modified1} of the main text, it is important to choose local excitation operators $\{ P_k^{(i)} \}$ so that they reproduce excitations by intra-subsystem terms in addition to those by inter-subsystem terms.
While we choose Pauli operators on the whole subsystem as such excitation operators in the main text, we can also pick up the following local basis set instead:
\begin{equation} \label{eq: multistate basis}
    \bigcup_{m=1}^M \{ P_k^{(i)} \ket{\psi_m}_{i} \}_{k=1}^K,
\end{equation}
with the original choice of $\{ P_k^{(i)} \}$ (local operators relevant to the inter-subsystem interactions).
The low-energy excited states of each subsystem, $\ket{\psi_m}_i$, can be obtained by using VQE-based algorithms for excited states such as  SSVQE~\cite{nakanishi2018subspace} instead of the usual VQE in Step 1, with a $N_\mr{qubit}$-qubit device. The dimension of the local Hilbert space is $M \times K$, typically smaller than $2^{N_\mr{qubit}}$.

Reference~\cite{fujii2020deep} reported that the Deep VQE based on this choice with $M>1, K=1$ fails to capture the ground state of the whole system, since it neglects the local excitations coming from inter-subsystem interactions. However, if we consider Deep VQE with $M>1$ and $K>1$, it is expected to well describe both the ground state and the low-energy eigenstates since it captures excitations from both inter-subsystem and intra-subsystem terms, just as the modified Deep VQE in the main text.
One of the possible problems of this version of Deep VQE compared to the one in the main text lies in calculating matrix elements of the effective Hamiltonian in Step 2: we should compute off-diagonal elements ${}_i\! \braket{\psi_m | P_k^{(i),\dagger} P_l^{(i)} | \psi_n}_i$
 and ${}_i\! \braket{\psi_m | P_k^{(i),\dagger} A P_l^{(i)} | \psi_n}_i$ ($A=H_i, V_i^\alpha, W_i^\alpha$) for $m \neq n$.
 We should choose a proper method for calculating the excited states of the subsystem in Step 1 of Deep VQE that enables us to evaluate those off-diagonal elements easily.
 The algorithms SSVQE and MCVQE satisfy this requirement while VQD dose not, demanding auxiliary qubits and/or the increase of the circuit depth to evaluate the matrix elements. 
 
 Thus, the choice of the local basis described in the main text is more preferable for calculating low-energy eigenstates in that we can use any kind of VQE-based algorithms in Step 1 of Deep VQE.
 On the other hand, the multi-state local basis like Eq.~\eqref{eq: multistate basis} can be suitable compared to the one in the main text when the intra-subsystem Hamiltonian is non-local enough to break the validity of QSE in each subsystem (see discussion in Sec.~\ref{subsec:modified1}).
 We also note that, when the gap above the subsystem ground state $\ket{\psi_0}_i$ is small enough, we may have to consider the multi-state local basis to capture the low-energy states for the whole system.

\section{\label{Asec:construction} Proof of Proposition in Sec. \ref{subsec:modified2}}
We derive the proposition in Section \ref{subsec:modified2}.
First, we clarify the setup for the construction of the effective qubit model and the notation. We consider an effective Hamiltonian
\begin{equation}\label{Aeq: hamiltoniantilde}
\tilde{H} = \left. H \right|_{\tilde{\mathcal{H}} } = \sum_{i=1}^{N_{\text{sub}}} \tilde{H}_i + \sum_{i \neq j}^{N_{\text{sub}}} \sum_\alpha \nu_{ij}^\alpha \tilde{V}_i^\alpha \otimes \tilde{W}_j^\alpha,
\end{equation}
which is defined on the $K^{N_{\text{sub}}}$-dimensional Hilbert space $\tilde{\mathcal{H}}$ as a result of choosing the local basis (see Step 2 in Section \ref{sec:preli}). We can regard the Hamiltonian $\tilde{H}$ as that on a lattice $\Lambda=\{1,\hdots,i,\hdots,N_{\text{sub}}\}$ where each subsystem $i \in \Lambda$ has $K$ levels. We define $N_{\text{eff}}$ by the smallest integer that exceeds $\log_2 K$, and we introduce auxiliary dimensions by
\begin{eqnarray}
\tilde{H}_{\text{eff}} &=& \sum_{i=1}^{N_{\text{sub}}} \tilde{H}_{i,\text{eff}} + \sum_{i \neq j}^{N_{\text{sub}}} \sum_\alpha \nu_{ij}^\alpha \tilde{V}_{i,\text{eff}}^\alpha \otimes \tilde{W}_{j,\text{eff}}^\alpha, \label{Aeq: hamiltonian_eff_1} \\
\tilde{H}_{i,\text{eff}} &=& \tilde{H}_{i} \oplus J_i, \label{Aeq: hamiltonian_eff_2} \\ \tilde{V}_{i,\text{eff}}^\alpha &=& \tilde{V}_{i}^\alpha \oplus 0_{M}, \quad \tilde{W}_{j,\text{eff}}^\alpha = \tilde{W}_{j}^\alpha \oplus 0_{M}, \label{Aeq: hamiltonian_eff_3}
\end{eqnarray}
with $M=2^{N_{\text{eff}}}-K$. As discussed in Section \ref{sec:preli}, this insertion of auxiliary dimension generates additional meaningless eigenvalues giving strong limitation on variational quantum circuits. We derive the way to construct $M \times M$ matrices $\{ J_i \}_{i=1}^{N_{\text{sub}}}$ so that the Hamiltonians $\tilde{H}$ and $\tilde{H}_{\text{eff}}$ can have the same low-energy spectrum.

We introduce some notations for matrices. For a finite-dimensional hermitian matrix $A$, we describe its eigenvalues by $E_0(A),E_1(A),\hdots,$ with $E_n(A)\leq E_{n+1}(A)$ and the set of the eigenvalues by $\text{Spec}(A)=\{ E_n(A) \}_{n=0}^{\mathrm{dim}(A)-1}$. When $H$ is a Hamiltonian defined on a lattice of subsystems $\Lambda=\{1,\hdots,i,\hdots,N_{\text{sub}}\}$, we can always decompose it as 
\begin{equation}
    H = \sum_{X \subset \Lambda} h_{X,\Delta},
\end{equation}
where $h_{X,\Delta}$ is a hermitian operator nontrivially acting just on a domain $X \subset \Lambda$. For instance, two-body interactions are represented by a series of $h_{X,\Delta}$ with $|X|=2$. We note that the way of the decomposition is not unique, and hence we designate its choice by the subscript $\Delta$.  Then, we define $H(D)$ for a certain domain $D \subset \Lambda$ by
\begin{equation}\label{Aeq:H_D_def}
    H_{\Delta}(D) = \sum_{X: \, X \cap D \neq \phi} h_{X,\Delta},
\end{equation}
which represents the part of $H$ nontrivially acting on some sites in $D$. When we choose a domain $D$ as the $i$-th subsystem, it becomes
\begin{equation}
    H_{\Delta}(D=\{ i \}) = \sum_{X: \, X \ni i} h_{X,\Delta}.
\end{equation}
Then, we define the extensiveness of Hamiltonian $H$ for the $i$-th subsystem by
\begin{equation}\label{Aeq:extensiveness_def}
    e_{H,\Delta}(i) \equiv \sum_{X: \, X \ni i} || h_{X,\Delta} ||_{\mr{op}},
\end{equation}
in which $|| \quad ||_{\mr{op}}$ denotes the operator norm. The extensiveness $e_{H,\Delta}(i)$ represents the maximal energy on the $i$-th subsystem under the Hamiltonian $H$.
Note that $e_{H,\Delta}(i)$ depends on the decomposition of the Hamiltonian $\Delta$. When we designate the decomposition by Eq. (\ref{Aeq: hamiltoniantilde}), the extensiveness becomes
\begin{equation}\label{Aeq: extensiveness_decomposition}
    e_{\tilde{H}}(i) = ||\tilde{H}_i||_\mr{op} + \sum_\alpha \sum_{j,k: \, \{j,k\} \ni i} |\nu_{jk}^\alpha| \cdot ||\tilde{V}_j^\alpha||_\mr{op} \cdot ||\tilde{W}_k^\alpha||_\mr{op}
\end{equation}
However, the choice is not essential in the discussion below, and hence we omit the subscript $\Delta$ like $h_X$, $H(D)$, and $e_H(i)$ in the following.

Before proving the proposition in the main text, we show the relation between eigenvalues of $\tilde{H}$ and those of $\tilde{H}_{\text{eff}}$. The result is summarized as follows.

\begin{proposition}\label{Aprop:1}
We introduce auxiliary dimension by Eqs. (\ref{Aeq: hamiltonian_eff_1})-(\ref{Aeq: hamiltonian_eff_3}). 
For a domain $D \subset \Lambda$, we define a $(K^{N_{\mr{sub}}-|D|}\times M^{|D|})$-dimensional Hilbert space $\tilde{\mathcal{H}}_D$, where subsystems out of $D$ and those in $D$ have $K$ and $M$ degrees of freedom, respectively. Then, the set of eigenvalues of $\tilde{H}_{\mr{eff}}$ is decomposed as follows:
\begin{equation}\label{Aeq:spec}
    \mr{Spec}(\tilde{H}_{\mr{eff}}) = \bigcup_{D \subset \Lambda} \mr{Spec}\left( \left[ \tilde{H} - \tilde{H}(D) \right]_{\tilde{H}_D} + \sum_{i \in D} J_i \right),
\end{equation}
in which we count the eigenvalues with including their degeneracy. Here, $[ \tilde{H} - \tilde{H}(D) ]_{\tilde{H}_D}$, described by
\begin{equation}\label{Aeq:H_H_D}
   \left[ \tilde{H} - \tilde{H}(D) \right]_{\tilde{\mathcal{H}}_D} = \sum_{i \notin D} \tilde{H}_i + \sum_\alpha \sum_{i \neq j: \, i,j \notin D} \nu_{ij}^\alpha \tilde{V}_i^\alpha \otimes \tilde{W}_j^\alpha,
\end{equation}
and $\sum_{i \in D} J_i$ are considered as $(K^{N_{\mr{sub}}-|D|}\times M^{|D|})$-dimensional matrices defined on $\tilde{\mathcal{H}}_D$.
\end{proposition}

\textit{Proof}--- Let $Q_i$ denote a projection operator to the auxiliary $M$-dimensional subspace on the $i$-th subsystem:
\begin{equation}
    Q_i \equiv \left( 0_{K} \oplus I_{M} \right)_i.
\end{equation}
Since $[Q_i, \tilde{H}_{\mr{eff}}]=0$ and $[Q_i,Q_j]=0$ are satisfied for any $i, j \in \Lambda$, the Hamiltonian $\tilde{H}_{\mr{eff}}$ can be block-diagonalized by adopting the eigenstates of $Q_i$ as the basis. Then, each block becomes the $(K^{|D|}\times M^{N_{\mr{sub}}-|D|})$-dimensional Hilbert space $\tilde{\mathcal{H}}_D$ since the projection to $\tilde{\mathcal{H}}_D$ is 
\begin{equation}
    P_{\mathcal{H}_D} =  \prod_{i \notin D} (1-Q_i) \cdot \prod_{i \in D} Q_i.
\end{equation}
As a result, the Hamiltonian $\tilde{H}_\mr{eff}$ on the extended $(K+M)^{N_\mr{sub}}$-dimensional Hilbert space $\tilde{\mathcal{H}}_\mr{eff}$ is decomposed as
\begin{equation}
    \tilde{H}_\mr{eff} = \bigoplus_{D \subset \Lambda} \left. \tilde{H}_\mr{eff} \right|_{\tilde{\mathcal{H}}_D},
\end{equation}
where $\left. \tilde{H}_\mr{eff} \right|_{\tilde{\mathcal{H}}_D}$ is obtained by projecting $\tilde{H}_\mr{eff}$ to $\tilde{\mathcal{H}}_D$. When we consider the simplest case $D=\{i \}$, the restricted Hamiltonian is given by
\begin{equation}\label{Aeq:h_restrict_D}
    \left. \tilde{H}_\mr{eff} \right|_{\tilde{\mathcal{H}}_D} = \sum_{j: \, j\neq i} \tilde{H}_j + J_i + \sum_\alpha \sum_{j,k: \, j,k \neq i} \nu_{jk}^\alpha \tilde{V}_j^\alpha \otimes \tilde{W}_k^\alpha.
\end{equation}
since $\tilde{H}_i$, $\tilde{V}_i^\alpha$, and $\tilde{W}_i^\alpha$ are respectively replaced to $J_i$, $0_M$, and $0_M$ by the projection $P_{\tilde{\mathcal{H}}_D}$.
We note that omitted identity operators in Eq. (\ref{Aeq:h_restrict_D}) are properly chosen from $I_K$ or $I_M$ to give a matrix on $\tilde{\mathcal{H}}_D$ here.
In a similar way, we can obtain those for a generic domain $D \subset \Lambda$ as
\begin{eqnarray}
    \left. \tilde{H}_\mr{eff} \right|_{\tilde{\mathcal{H}}_D} &=& \sum_{i \notin D} \tilde{H}_i + \sum_\alpha \sum_{i \neq j: \, i,j \notin D} \nu_{ij}^\alpha \tilde{V}_i^\alpha \otimes \tilde{W}_j^\alpha + \sum_{i \in D} J_i \nonumber \\
    &\equiv&  \left[ \tilde{H} - \tilde{H}(D) \right]_{\tilde{H}_D} + \sum_{i \in D} J_i.
\end{eqnarray}
This immediately results in Eq. (\ref{Aeq:spec}). $\quad \square$

When we choose $D=\phi$ in Prop. \ref{Aprop:1}, $[\tilde{H}-\tilde{H}(D)]_{\tilde{\mathcal{H}}_D}+\sum_{i \in D} J_i$ in Eq. (\ref{Aeq:spec}) is equivalent to the effective Hamiltonian $\tilde{H}$. Thus, all the eigenvalues of $\tilde{H}$ are embedded in those of $\tilde{H}_\mr{eff}$, though meaningless eigenvalues corresponding to $D \neq \phi$ appear. 

\begin{proposition}\label{Aprop:2}
Let $\tilde{H}_\mr{eff}$ be obtained from $\tilde{H}$ by means of the insertion of auxiliary dimensions described in Eqs. (\ref{Aeq: hamiltonian_eff_1})-(\ref{Aeq: hamiltonian_eff_3}). When we choose the $M \times M$ matrices $\{ J_i \}_{i=1}^{N_{\mr{sub}}}$ by
\begin{equation}\label{Aeq:J_bound}
    J_i = \lambda_i I_M, \quad \lambda_i > e_{\tilde{H}}(i) \text{ : extensiveness}
\end{equation}
for each subsystem $i \in \Lambda$, the ground-state energy of $\tilde{H}$ is equal to that of $\tilde{H}_\mr{eff}$, that is,
\begin{equation}\label{Aeq:ground_energy}
    E_0 (\tilde{H}) = E_0 (\tilde{H}_\mr{eff})
\end{equation}
is satisfied.
\end{proposition}

\textit{Proof.}---From the result of Prop. \ref{Aprop:1}, it is sufficient to show that
\begin{equation}\label{Aeq:goal_prop2}
    E_0 (\tilde{H}) < E_0 \left( \left[ \tilde{H}-\tilde{H}(D) \right]_{\tilde{\mathcal{H}}_D} + \sum_{i \in D} J_i \right)
\end{equation}
is satisfied for any domain $D \neq \phi$ when $J_i$ is designated by Eq. (\ref{Aeq:J_bound}). With the usage of Eq. (\ref{Aeq:J_bound}), we obtain
\begin{eqnarray}
[\text{r.h.s of Eq. (\ref{Aeq:goal_prop2})}]
&=&
E_0 \left( \left[ \tilde{H}-\tilde{H}(D) \right]_{\tilde{\mathcal{H}}_D} \right) + \sum_{i \in D} \lambda_i \nonumber \\
&>& E_0 \left( \left[ \tilde{H}-\tilde{H}(D) \right]_{\tilde{\mathcal{H}}_D} \right) + \sum_{i \in D} e_{\tilde{H}}(i). \nonumber
\end{eqnarray}
When we explicitly write the omitted identity operators in Eq. (\ref{Aeq:H_H_D}), $[ \tilde{H}-\tilde{H}(D) ]_{\tilde{\mathcal{H}}_D}$ becomes
\begin{equation}
    \left[ \sum_{i \notin D} \tilde{H}_i + \sum_\alpha \sum_{i \neq j: \, i,j \notin D} \nu_{ij}^\alpha \tilde{V}_i^\alpha \otimes \tilde{W}_j^\alpha \right] \otimes 
    \left[ \bigotimes_{i \in D} (I_M)_i \right].
\end{equation}
If we replace $\bigotimes_{i \in D} (I_M)_i$ by $\bigotimes_{i \in D} (I_K)_i$, we obtain $\tilde{H}-\tilde{H}(D)$, represented by a $K^{N_\mr{sub}}$-dimensional matrix. This results in the same ground-state energy,
\begin{equation}\label{Aeq:E0_same}
    E_0 \left( \left[ \tilde{H}-\tilde{H}(D) \right]_{\tilde{\mathcal{H}}_D} \right) = E_0 \left(  \tilde{H}-\tilde{H}(D)  \right),
\end{equation}
although their ground states have different degeneracy. Next, we exploit the following theorem for any hermitian operators $H$ and $H^\prime$ with the same finite dimension (see Section III. 2 in Ref. \cite{bhatia1997matrix}):
\begin{equation}
    | E_m (H) - E_m (H^\prime) | \leq || H - H^\prime ||_\mr{op}
\end{equation}
for any integer $m \geq 0$. Then we obtain
\begin{equation}
E_0 \left( \tilde{H} - \tilde{H}(D) \right) \geq E_0 (\tilde{H}) - || \tilde{H}(D) ||_\mr{op}.
\end{equation}
From the definition of $\tilde{H}(D)$ [see Eq. (\ref{Aeq:H_D_def})] and the extensiveness [see Eq. (\ref{Aeq:extensiveness_def})], we arrive at the following inequality:
\begin{eqnarray}
|| H(D) ||_\mr{op} &\leq& \sum_{X: \, X \cap D \neq \phi} || h_X ||_\mr{op} \nonumber \\
&\leq& \sum_{i \in D} \sum_{X: \, X \ni i} || h_X ||_\mr{op} 
= \sum_{i \in D} e_H (i) \label{Aeq:opnorm_bound}
\end{eqnarray}
for any Hamiltonian $H$ on the lattice $\Lambda$. As a result, we obtain
\begin{eqnarray}
[\text{r.h.s of Eq. (\ref{Aeq:goal_prop2})}]
&>& E_0 (\tilde{H}) - || \tilde{H}(D) ||_\mr{op} + \sum_{i \in D} e_{\tilde{H}}(i) \nonumber \\
&\geq& E_0 (\tilde{H}) \label{Aeq:goal_process2}
\end{eqnarray}
for $D \neq \phi$. Under this inequality, $\min \{ \mr{Spec} (\tilde{H}_\mr{eff}) \}$ is given by $E_0(\tilde{H})$, indicating Eq. (\ref{Aeq:ground_energy}). $\quad \square$

As discussed in the main text, this proposition is used for the modified way to construct the effective Hamiltonian $\tilde{H}_\mr{eff}$, which properly gives the ground state with arbitrary variational quantum states.
In order to evaluate low-energy eigenstates, Prop. \ref{Aprop:2} is extended to the following proposition, which is referred to as {\bf Proposition} in Sec.~\ref{subsec:modified2} of the main text.

\begin{proposition}\label{Aprop:3}
We consider the same setup as the one in Prop. \ref{Aprop:2}. For a certain integer $n \geq 0$, let us choose the $M \times M$ matrices $\{ J_i \}_{i=1}^{N_{\mr{sub}}}$ by
\begin{equation}\label{Aeq:J_bound_2}
    J_i = \lambda_i I_M, \quad \lambda_i > e_{\tilde{H}}(i) + E_n(\tilde{H})-E_0(\tilde{H}), \quad^\forall i \in \Lambda.
\end{equation}
Then, the low-energy eigenvalues of $\tilde{H}$ are equal to those of $\tilde{H}_\mr{eff}$ respectively up to the $n$-th level, that is,
\begin{equation}\label{Aeq:low_energy}
    E_m (\tilde{H}) = E_m (\tilde{H}_\mr{eff}), \quad m=0,1,\hdots,n
\end{equation}
is satisfied.
\end{proposition}

\textit{Proof.}---We prove this in the same way for Prop. \ref{Aprop:2}. From the result of Prop. \ref{Aprop:2}, we show that
\begin{equation}\label{Aeq:goal_prop3}
    E_n (\tilde{H}) < E_0 \left( \left[ \tilde{H}-\tilde{H}(D) \right]_{\tilde{\mathcal{H}}_D}+ \sum_{i \in D} J_i \right)
\end{equation}
is satisfied for any domain $D \neq \phi$ when $J_i$ is designated by Eq. (\ref{Aeq:J_bound_2}). By the same calculation as Eq. (\ref{Aeq:goal_process2}), we obtain
\begin{eqnarray}
[\text{r.h.s of Eq. (\ref{Aeq:goal_prop3})}]
&>& E_0 (\tilde{H}) - || \tilde{H}(D) ||_\mr{op} + \sum_{i \in D} \lambda_i \nonumber \\
&>& E_0 (\tilde{H}) + |D| \{ E_n (\tilde{H}) - E_0 (\tilde{H}) \} \nonumber \\
&\geq& E_n (\tilde{H})
\end{eqnarray}
for $D \neq \phi$. We use $|D| \geq 1$ under $D \neq \phi$. Under this inequality and Eq. (\ref{Aeq:spec}), Eq. (\ref{Aeq:low_energy}) is satisfied. $\quad \square$

Note that we should know the gap $E_n (\tilde{H}) - E_0 (\tilde{H})$ in advance when we construct $\tilde{H}_\mr{eff}$ following Eq. (\ref{Aeq:J_bound_2}) different from the case when evaluating the ground state. In practice, we know typical energy scale of the gap by some other methods in advance, and can properly choose $J_i$ as discussed in the main text. From the mathematical point of view, we can construct $\tilde{H}_\mr{eff}$ without the knowledge of the gap  $E_n (\tilde{H}) - E_0 (\tilde{H})$ in advance even when we evaluate excited states of $\tilde{H}$, by the following proposition.

\begin{proposition}\label{Aprop:4}
We consider the same setup as the one in Prop. \ref{Aprop:2}. Let us choose the $M \times M$ matrices $\{ J_i \}_{i=1}^{N_{\mr{sub}}}$ by
\begin{equation}\label{Aeq:J_bound_3}
    J_i = \lambda_i I_M, \quad \lambda_i > e_{\tilde{H}}(i) + 2 \sum_{j \in \Lambda} e_{\tilde{H}}(j), \quad^\forall i \in \Lambda.
\end{equation}
Then, all of the eigenvalues of $\tilde{H}$ appear in the beginning of $\mr{Spec}(\tilde{H}_\mr{eff})$, that is,
\begin{equation}\label{Aeq:full_energy}
    E_m (\tilde{H}) = E_m (\tilde{H}_\mr{eff}), \quad m=0,1,\hdots,K^{N_\mr{sub}}-1
\end{equation}
is satisfied.
\end{proposition}

\textit{Proof.}---Under the choice of Eq. (\ref{Aeq:J_bound_3}),
\begin{eqnarray}
\lambda_i &>& e_{\tilde{H}}(i) + 2 || \tilde{H} ||_\mr{op} \nonumber \\
&\geq& e_{\tilde{H}}(i) + E_{K^{N_\mr{sub}}-1}(\tilde{H})-E_0(\tilde{H})
\end{eqnarray}
is satisfied. We use the result of Prop. \ref{Aprop:3} with $n=K^{N_\mr{sub}}-1$, and then we obtain Eq. (\ref{Aeq:full_energy}). $\quad \square$

We note that, although Prop. \ref{Aprop:4} provides a mathematically right way to evaluate excited states without the knowledge of the gap in advance, it is not practical compared to the one exploiting Prop. \ref{Aprop:3}.
When we choose $J_i$ by Eq. (\ref{Aeq:J_bound_3}), its energy scale becomes linear in the system size $N_{\mr{sub}} \times N_\mr{qubit}$.
When considering a large system, $J_i$ becomes dominant in the effective Hamiltonian $\tilde{H}_{i,\mr{eff}}$ [see Eqs. (\ref{Aeq: hamiltonian_eff_1})-(\ref{Aeq: hamiltonian_eff_3})] compared to the original one $\tilde{H}_i$. Then, errors originating from $J_i$ buries the information of $\tilde{H}_i$, which is genuinely of interest, and hence the insertion of auxiliary dimensions based on Prop. \ref{Aprop:4} is not practical, though it is mathematically correct.

We finally discuss the generalization of our results. In the above discussion, we designate the decomposition of the Hamiltonian $\tilde{H}$ by Eq. (\ref{Aeq: hamiltoniantilde}). When we choose another decomposition, we obtain a different extended Hamiltonian $\tilde{H}_\mr{eff}$ [Eqs. (\ref{Aeq: hamiltonian_eff_1})-(\ref{Aeq: hamiltonian_eff_3})] and a different extensiveness $e_{\tilde{H}}(i)$ [Eq. (\ref{Aeq: extensiveness_decomposition})]. However, the same results, Props. \ref{Aprop:1}-\ref{Aprop:4}, are valid due to the relations Eqs. (\ref{Aeq:E0_same}) and (\ref{Aeq:opnorm_bound}). Thus, the decomposition is not essential. As well, we can consider inter-subsystem interactions which simultaneously involve more than three subsystems. In such cases, we introduce auxiliary dimensions for intra-subsystem terms by a nonzero matrix $J_i$ and for inter-subsystem terms by a zero matrix $0_{2^{N_\mr{eff}}-K}$ as well as Eqs. (\ref{Aeq: hamiltonian_eff_1})-(\ref{Aeq: hamiltonian_eff_3}). The extensiveness $e_H(i)$ is given by the same definition Eq. (\ref{Aeq:extensiveness_def}). Then, using the relations Eqs. (\ref{Aeq:E0_same}) and (\ref{Aeq:opnorm_bound}), Props. \ref{Aprop:1}-\ref{Aprop:4} are valid in the cases with interactions involving more than three subsystems.

\section{Details of numerical simulations \label{Aseq:ansatz}}

\begin{figure}
\begin{center}
    \includegraphics[height=4cm, width=9cm]{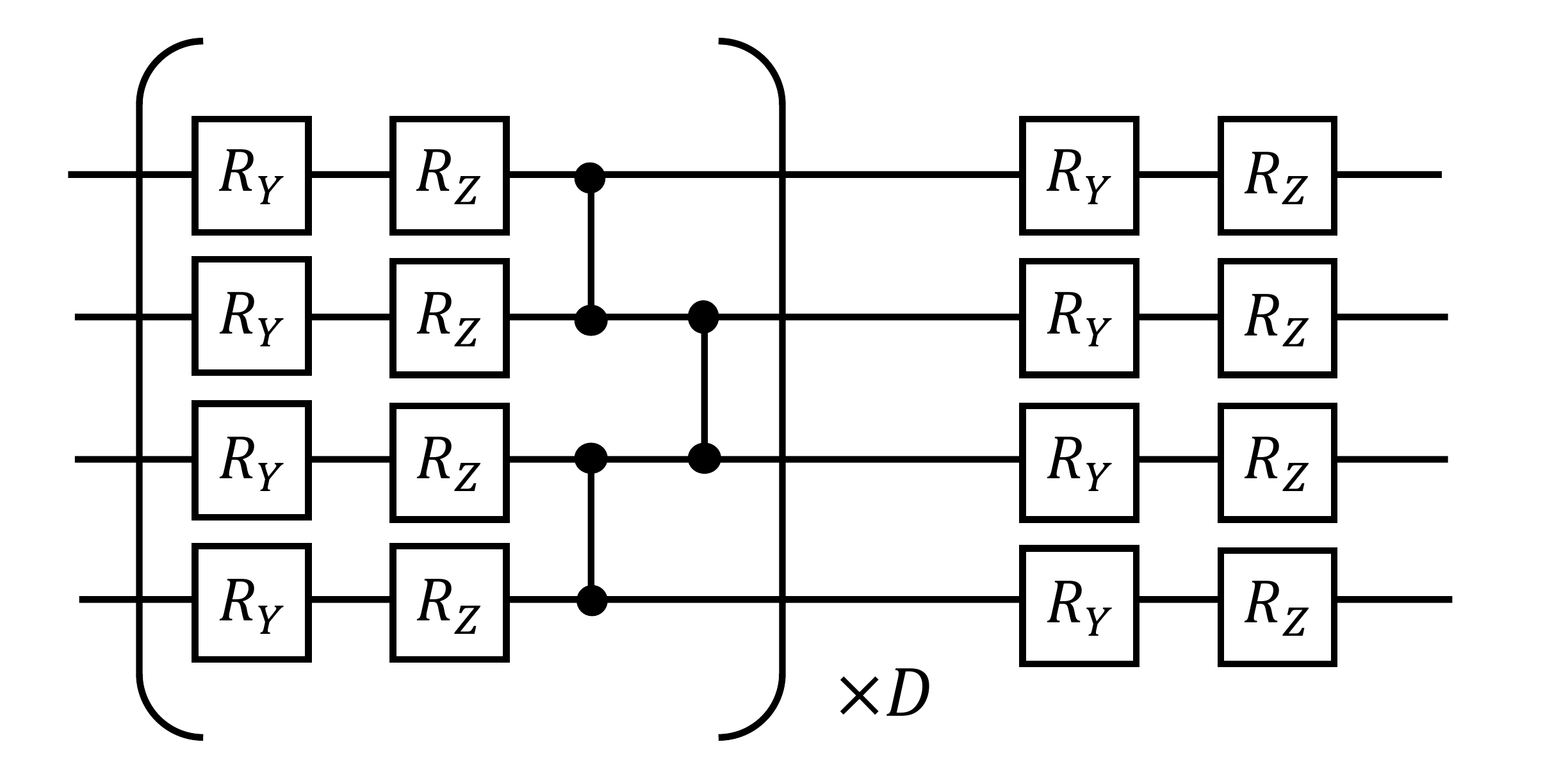}
    \caption{Variational quantum circuit based on Hardware-efficient ansatz. We prepare the reference states by $\ket{0\hdots 00}$ for the VQE and $\ket{0\hdots 00},\ket{0\hdots 01}$ for the SSVQE. } 
    \label{fig:ansatz}
 \end{center}
 \end{figure}

Here, we describe the details of numerical simulations in the main text, especially focusing on the variational quantum circuits and the cost function.
First, we identify the variational quantum circuits in VQE in Step 1 and Step 3 of Deep VQE.
We employ the hardware-efficient-type ansatz~\cite{kandala2017hardware,Mitarai2018quantum}, where the parametric circuit with the depth $D$ is described by
\begin{eqnarray}
    U(\vec{\theta}) &=& U_1 (\vec{\theta}_{D+1},\vec{\theta^\prime}_{D+1}) \prod_{d=1}^D \left[ U_2 \cdot U_1 (\vec{\theta}_{d},\vec{\theta^\prime}_{d}) \right], \qquad \\
    U_1 (\vec{\theta}_{d},\vec{\theta^\prime}_{d}) &=& \prod_{n=1}^N \left[ R_{Z,n}(\theta^\prime_{d,n}) \cdot R_{Y,n}(\theta_{d,n}) \right], \\
    U_2 &=& \prod_{n=1}^{N-1} \mr{CZ}_{n,n+1},
\end{eqnarray}
for a $N$-qubit system (see Fig. \ref{fig:ansatz}). Here, $R_{Y,n}(\theta)$ [$R_{Z,n}(\theta)$] represents rotation around $Y$-axis [$Z$-axis] with an angle $\theta$ on the $n$-th qubit, and $\mr{CZ}_{n,m}$ is a controlled-Z gate on the $n$-th and $m$-th qubits. The parameter set $\vec{\theta}$ is composed of $\{ \vec{\theta}_d \}_{d=1}^{D+1}$ and $\{ \vec{\theta}^\prime_d \}_{d=1}^{D+1}$, whose initial values are chosen uniformly from $[0,2\pi)^{2N(D+1)}$ at random.
When we evaluate the ground state in Step 1, we employ the cost function $\braket{0 \hdots 0 | U(\vec{\theta})^\dagger H_i U(\vec{\theta}) | 0 \hdots 0}$ for each subsystem-Hamiltonian $H_i$.
For evaluating the first-excited states in Step 3 of Deep VQE, we employ SSVQE~\cite{nakanishi2018subspace}, in which the cost function is given by
\begin{eqnarray*}
 && w_0 \braket{0 \hdots 00 | U(\vec{\theta})^\dagger \tilde{H}_\mr{eff} U(\vec{\theta}) | 0 \hdots 00} \\
    &\qquad& \qquad \qquad \qquad + w_1 \braket{0 \hdots 01 | U(\vec{\theta})^\dagger \tilde{H}_\mr{eff} U(\vec{\theta}) | 0 \hdots 01}
\end{eqnarray*}
with $w_0>w_1>0$. After the optimization, we can obtain the approximate ground (first-excited) state by $U(\vec{\theta}^\ast)\ket{0\hdots 00}$ ($U(\vec{\theta}^\ast)\ket{0\hdots 01}$). We employ the BFGS optimizer for optimizing the circuit parameters with numerical differentiation. 

For the spin chain in Section \ref{subsec:spinchain}, we set the depth $D$ by $10,15,20$ for $4,6,8$-qubit systems in Step 1, and by $15,20,25$ for $4,6,8$-qubit systems in Step 3, respectively.
The set of weights for SSVQE is $(w_0,w_1)=(2,1)$.
For the periodic hydrogen chain in Section \ref{sec:chemistry}, we employ the depth $D=20$ in Step 1 and $D=80$ in Step 3, and choose the weights by  $(w_0,w_1)=(7,2)$.

\section{\label{Asec: DeepVQE and QSE} Relation between Deep VQE with higher-order excitation and QSE results}
In Section \ref{subsec:hydrogen_results}, the local basis by the double-particle excitation $\mathcal{W}_\mr{d}$ gives better low-energy eigenstates than the one by single-particle excitation $\mathcal{W}_\mr{s}$. Here, we discuss the role of higher-order excitations in Deep VQE. Concretely, we show that Deep VQE with higher-order excitations overwhelms the combination of QSE and Deep VQE with lower-order excitations---calculating the ground state by Deep VQE with lower-order excitations and performing QSE on the obtained ground state with lower-order excitations. This gives the advantage of considering higher-order excitations in Deep VQE.

Let us compare the above two methods by the active subspaces in them.  For simplicity, we consider single-particle (the first-order) and double-particle (the second-order) excitations. First, we focus on the combination of QSE and Deep VQE with single-particle excitation $\mathcal{W}_\mr{s}$ [see Eq. (\ref{eq:single_excitation})]. The Deep VQE with $\mathcal{W}_\mr{s}$ searches low-lying states from the subspace $\tilde{\mathcal{H}}_\mr{s}$, spanned by
\begin{equation}
     \ket{\Psi_0}, \, R_i \ket{\Psi_0}, \, R_i R_j \ket{\Psi_0}, \, R_i R_j R_k \ket{\Psi_0}, \hdots.
 \end{equation}
 Here, $R_i$ is given by $c_i^\prime$ or $c_i^{\prime \dagger}$ and the indices $i,j,k,\hdots$ represent certain sites of the whole system belonging to different subsystems each other. Let $\ket{\Psi_0^\mr{s}(\text{Deep VQE})} \in \tilde{\mathcal{H}}_\mr{s}$ denote the approximate ground state by the Deep VQE with $\mathcal{W}_\mr{s}$. The QSE with the reference state $\ket{\Psi_0^\mr{s}(\text{Deep VQE})}$ and the single-particle excitations in the whole-system $R_i=c_i^\prime \,  \text{or} \, c_i^{\prime \dagger}$ searches low-energy eigenstates from the subspace
\begin{equation}
    \tilde{\mathcal{H}}_\mr{s}^\mr{QSE} = \mr{span} \left( \{ \ket{\Psi_0^\mr{s}(\text{Deep VQE})} , \, R_i \ket{\Psi_0^\mr{s}(\text{Deep VQE})} \} \right).
 \end{equation}

On the other hand, we next focus on the Deep VQE result with the complete double-particle excitations,
\begin{equation}
     \mathcal{W}_\mr{d}^\prime = \mathcal{W}_\mr{s} \cup \{ c_i^{\prime \dagger} c_j^\prime, \, c_i^\prime c_j^\prime, \, c_i^{\prime \dagger} c_j^{\prime \dagger} \, | \, i,j \in \Lambda_e \},
 \end{equation}
By construction, the restricted subspace $\tilde{\mathcal{H}}_\mr{d}^\prime$, which is active in the Deep VQE with $\mathcal{W}_\mr{d}^\prime$, includes not only states in $\tilde{\mathcal{H}}_\mr{s}$ but also
\begin{equation}
     R_i \ket{\Psi} \quad \text{for any state} \quad \ket{\Psi} \in \mathcal{\tilde{H}}_\mr{s},
 \end{equation}
 with single-particle excitations $R_i=c_i^\prime \, \text{or} \, c_i^{\prime \dagger}$. Considering the fact $\ket{\Psi_0^\mr{s}(\text{Deep VQE})} \in \tilde{\mathcal{H}}_\mr{s}$, we obtain
 \begin{equation}
     \tilde{\mathcal{H}}_\mr{s}^\mr{QSE} \subset \tilde{\mathcal{H}}_\mr{d}^\prime.
 \end{equation}
This means that the Deep VQE results with double-particle excitations are more accurate than the QSE results on the Deep VQE with single-particle excitations. We stress that this statement on the difference of the accuracy between single- and double-particle excitations is more strict than the one by simply comparing the Deep VQE results with $\mathcal{W}_\mr{s}$ and $\mathcal{W}_\mr{d}^\prime$ from the relation $\tilde{\mathcal{H}_\mr{s}} \subset \tilde{\mathcal{H}}_\mr{d}^\prime$.

In general, the Deep VQE results with local excitations up to $n$-th order always give better low-lying energy eigenvalues than the results obtained by the QSE with up to $(n-m)$-th order excitations after performing the Deep VQE with up to $m$-th order excitations ($0 \leq m \leq n$). Although our calculation with $\mathcal{W}_\mr{d}$ in the main text [Eq. (\ref{eq:double_excitation})] does not strictly satisfy the above discussion due to the absence of $c_i^\prime c_j^\prime$ and $c_i^{\prime \dagger} c_j^{\prime \dagger}$, we expect that the restricted space by $\mathcal{W}_\mr{d}$, similar to $\tilde{\mathcal{H}}_\mr{d}^\prime$, explains the improved accuracy compared to the single-particle excitations $\mathcal{W}_\mr{s}$. For larger systems, we can exploit a complete set of higher-order excitations for constructing the local basis with keeping the reduction of qubits. In that case, we expect much improved accuracy of the Deep VQE results by higher-order excitations from their superiority to the combination of QSE and Deep VQE with lower-order excitations.
\end{document}